\documentclass[a4paper,11pt]{article}

\usepackage{jheppub} 
\usepackage[T1]{fontenc} 
\usepackage{bm}
\newcommand{\tr}{\,\mbox{tr}}
\newcommand{\sign}[1]{\,\mbox{sgn}\left({#1}\right)}

\definecolor{purple}{rgb}{0.8,0,0.6}
\definecolor{darkgreen}{rgb}{0.00,0.6,0.00}

\title{Wigner function and kinetic phenomena for chiral plasma in a strong magnetic field}

\author[a,b]{E. V. Gorbar,}
\author[c]{V. A. Miransky,}
\author[d,e,1]{I. A. Shovkovy\note{Corresponding author.}}
\author[c]{and P. O. Sukhachov}

\affiliation[a]{Department of Physics, Taras Shevchenko National Kiev University, Kiev, 03680, Ukraine}
\affiliation[b]{Bogolyubov Institute for Theoretical Physics, Kiev, 03680, Ukraine}
\affiliation[c]{Department of Applied Mathematics, Western University, London, Ontario N6A 5B7, Canada}
\affiliation[d]{College of Integrative Sciences and Arts, Arizona State University, Mesa, Arizona 85212, USA}
\affiliation[e]{Department of Physics, Arizona State University, Tempe, Arizona 85287, USA}

\emailAdd{gorbar@bitp.kiev.ua}
\emailAdd{vmiransk@uwo.ca}
\emailAdd{igor.shovkovy@asu.edu}
\emailAdd{psukhach@uwo.ca}

\abstract{
By using the exact solutions of the Weyl equation in a constant magnetic field,
the equal-time Wigner function for magnetized chiral plasma is derived. It is found that
the dependence of the Wigner function on the component of momentum along the magnetic
field is asymmetric and is correlated with the fermion chirality. Such a dependence is principal
for reproducing the correct chiral magnetic and chiral separation effects. In the lowest Landau
level approximation, the equation for the equal-time Wigner function in a strong magnetic field
is derived. By making use of this equation, it is found that the longitudinal collective modes in
a strong magnetic field are gapped plasmons whose gap is determined by the magnetic field.
Unlike the ordinary magnetic field, an axial one allows for the dispersion law of the collective
excitations asymmetric in the wave vector. The thermoelectric phenomena for chiral fermions
in strong magnetic and axial magnetic fields are studied and the corresponding transport
coefficients are calculated.
}

\begin{document}
\maketitle
\flushbottom

\section{Introduction}
\label{sec:Introduction}

Kinetic theory \cite{Landau:t10,Liboff} is the standard and surprisingly very efficient method of the
investigation of transport phenomena in various physical systems. Classical kinetic theory is widely
used to describe the fluid dynamics, electromagnetic collective excitations in plasmas, conductivity
in metals, etc. Quantum kinetic theory is indispensable in the study of transport phenomena in
material media at low temperature. Relativistic kinetic theory is relevant in the studies of the
primordial plasma of the early Universe \cite{Vallee,Durrer}, relativistic heavy-ion collisions
\cite{Kharzeev:2008-Nucl,Kharzeev:2016}, compact stars \cite{Kouveliotou:1999}, and the
recently discovered Dirac \cite{Borisenko,Neupane,Liu,Xiong,Li-Wang:2015,Li-He:2015,Li} and Weyl
\cite{Weng-Fang:2015,Qian,Huang:2015eia,Bian,Huang:2015Nature,Zhang:2016,Cava,Belopolski}
materials whose low-energy excitations are described by the relativistic-like Dirac and Weyl
equations, respectively.

One of the qualitatively new key ingredients in the chiral plasma is the chiral charge,
whose conservation is violated only by the chiral quantum anomaly \cite{ABJ,ABJ-1}. Recently, it was shown that its dynamical evolution can be described using the framework of the kinetic theory.
The corresponding
version of the theory, i.e., the chiral kinetic theory was formulated in Refs.~\cite{Son,SonYamamoto,Stephanov}. This theory relies on the wave-packet semiclassical description of the anomalous
Hall effect \cite{Sinova} in metals which takes into account the Berry curvature effects \cite{Berry:1984,Xiao}. The latter are relevant
for the chiral kinetic theory because the Weyl nodes act as the sources and sinks of the Berry curvature. The chiral kinetic theory was derived
in the first order in the Planck constant $\hbar$ and its equations are linear in electric as well as magnetic fields. Such a
theory successfully incorporates the chiral anomaly \cite{ABJ,ABJ-1} and describes the chiral magnetic effect \cite{CME}. However, many physical
phenomena require the chiral kinetic theory accurate to higher orders in the electromagnetic field. The first step in this direction was done in
Refs.~\cite{Gao:2014,Gao:2015} where, by using the wave-packet semiclassical approach, the chiral kinetic equation valid to the second order in a
magnetic field was derived for fermions with a general band structure. In our recent paper \cite{Gorbar:2017cmw}, we provided the explicit
expressions for the chiral kinetic theory accurate to the second order in electromagnetic and axial or
pseudoelectromagnetic fields for a simple realization of relativistic matter in a Weyl material with a single pair of Weyl nodes.

Certainly, it would be very useful to formulate the equations of the chiral kinetic theory to all orders in electromagnetic fields. A
convenient starting point for their derivation is the equation of motion for the Wigner function \cite{Wigner:1932} (see also Refs.~\cite{Elze:1986,Vasak:1987,Elze:1989,Zachos,Polkovnikov}) in an
electromagnetic field. In a many-body system, the Wigner function is given by the Fourier transform of the two-point lesser Green's
function with respect to the relative coordinates (see, e.g., Ref.~\cite{Calzetta}). Therefore, similarly to the usual distribution function, the Wigner function
describes the dynamics in the phase space, albeit retaining all its quantum aspects.

The equation of motion for the Wigner function in an electromagnetic field is exact and mathematically equivalent to the Dirac or Weyl
equation. By solving this equation in the perturbation theory in electromagnetic field with the zero-order Wigner function proportional to a combination of
the Fermi--Dirac distribution functions, the standard equations of the chiral kinetic theory were derived in Ref.~\cite{Wang-kinetic}. On the other
hand, it is very well known that the Dirac or Weyl equation is exactly solvable in a constant magnetic field. This
suggests a possible means to analyse the kinetic phenomena for chiral fermions when the background magnetic field is taken into account
nonperturbatively and the corresponding Wigner function is found exactly. Further, this solution can be used to analyze perturbatively the effects
of a weak electric field, as well as inhomogeneous and time-dependent magnetic fields. This
provides the main motivation for the study performed in this paper. In addition, we would like to mention
also that such an approach could, in principle, be extended to the case of a constant background
electric field. However, unlike the magnetic field, the electric field does work. Therefore, even a
stationary solution would describe a nonequilibrium state, in which the pair production
can take place. Using the formalism of the equal-time Wigner function (known also as the Dirac-Heisenberg-Wigner
function), this problem was analyzed in Refs.~\cite{BialynickiBirula:1991tx,Gies:2010}. The rearrangement of the particle occupation numbers
makes the analysis more complicated, therefore, in this paper we consider only the case of
constant background magnetic and axial magnetic fields.

By making use of the equal-time Wigner function, in this study we also address the thermoelectric properties of
chiral plasma in the strong magnetic field limit. In weak fields, the corresponding anomalous transport was
studied in Refs.~\cite{Xiao-Niu:2006,Qin-Niu-Shi:2011} by taking into account the Berry phase effects.
Later, these thermoelectric properties were investigated in the framework of the chiral kinetic
theory for Dirac and Weyl materials in Refs.~\cite{Lundgren:2014hra,Sharma:2016-Weyl,Sharma:2016-Dirac,Tabert:2016,Chen-Fiete:2016}.
The authors of Ref.~\cite{Lundgren:2014hra} showed that the
thermoconductivity has the standard linear temperature dependence expected for a metallic system.
In the case where the temperature gradient and weak magnetic field are parallel, the longitudinal
thermoconductivity is quadratic in the magnetic field strength, which is somewhat similar to the
quadratic correction to the electric conductivity from the chiral anomaly. If the magnetic field is
perpendicular to the temperature gradient, then the magnetic field correction to the thermoconductivity
is also quadratic, albeit has a negative sign. In this study, we extend the analysis of the thermoelectric
properties of chiral matter to the case of strong background magnetic and axial magnetic fields.

The paper is organized as follows. The model is described in section~\ref{sec:model-Weyl}. The equal-time
Wigner function for chiral fermions in background constant magnetic and axial (or pseudo-) magnetic
fields is explicitly calculated in section~\ref{sec:Wigner-Weyl-magnetic}. In the same section we tested
the obtained Wigner function by studying the chiral magnetic and chiral separation effects, as well as
determining its weak magnetic field expansion. The equations of motion for the equal-time Wigner
function in the lowest Landau level approximation are given in section~\ref{sec:Wigner-Weyl-Eq}. The
collective excitations as well as the charge and heat transport in the longitudinal direction with respect to the
magnetic field are analyzed in sections~\ref{sec:Wigner-Weyl-LLL-plasmon} and
\ref{sec:Wigner-C-LLL-Termal}, respectively. The results are summarized in section~\ref{sec:summary}.
Some technical details of the derivation and analysis are collected in several appendices at the end
of the paper.

Throughout this paper, we set $\hbar=1$ and $c = 1$.

\section{Model}
\label{sec:model-Weyl}

The Weyl Hamiltonian for the right-handed $\lambda=+$ and left-handed $\lambda=-$ fermions in a constant magnetic field is given
by
\begin{equation}
H_{\lambda} = -i\lambda v_F (\bm{\sigma}\cdot\bm{\nabla})+\lambda v_Fe (\bm{\sigma}\cdot\mathbf{A}_{\lambda}) +\mu_{\lambda},
\label{model-Hamiltonian}
\end{equation}
where $v_F$ equals the velocity of light $c$ in the case of relativistic matter or the Fermi velocity of quasiparticles
in a Weyl semimetal.
Further, $\bm{\sigma}=\left(\sigma_x,\sigma_y,\sigma_z\right)$ are the Pauli matrices,
$\mathbf{A}_{\lambda}=\left(0,x_1 B_{\lambda},0\right)$ is a vector potential for an effective
magnetic field $\mathbf{B}_{\lambda}=\mathbf{B}+\lambda \mathbf{B}_{5}$ that points
in the $+z$ direction, and $\mu_{\lambda}=\mu+\lambda\mu_{5}$ is an effective
chemical potential, where $\mu$ is the fermion number chemical potential and $\mu_5$ is the chiral chemical potential. Here, we included the interaction of the chiral fermions with an axial magnetic (or, equivalently,
pseudomagnetic) field $\mathbf{B}_5$. Such a field interacts with the different sign depending on
the fermion chirality. While this field is typically absent in relativistic matter systems (except
for, possibly, in the primordial plasma of the early Universe before the electroweak transition)
it can be easily induced by mechanical strains in Weyl and Dirac semimetals
\cite{Zhou:2012ix,Zubkov:2015,Cortijo:2016yph,Cortijo:2016,Grushin-Vishwanath:2016,Pikulin:2016,Liu-Pikulin:2016}. The characteristic strengths of the pseudomagnetic field in Dirac and Weyl materials range
from about $B_5\approx0.3~\mbox{T}$, when a static torsion is applied to a nanowire of Cd$_3$As$_2$ \cite{Pikulin:2016}, to approximately
$B_5\approx15~\mbox{T}$, when a thin film of the same material is bent \cite{Liu-Pikulin:2016}.

The wave functions of Hamiltonian (\ref{model-Hamiltonian}) are derived in
appendix~\ref{sec:App-model-Weyl}. Their final expressions read
\begin{subequations}
\label{model-Bn0-WF-fin-n-all}
\begin{eqnarray}
\label{model-Bn0-WF-fin-n=0-1}
\psi_{n=0,p_2,p_3}(\mathbf{x}) &=& |eB_{\lambda}|^{1/4} e^{ip_3x_3 +ip_2x_2 }Y_0(\xi) \mathcal{P}_{s_B} \left(
                                                        \begin{array}{c}
                                                          0 \\
                                                          1 \\
                                                        \end{array}
                                                      \right),\\
\psi_{n>0,p_2,p_3}(\mathbf{x})  &=& |eB_{\lambda}|^{1/4} \sqrt{\frac{2v_F^2n |eB_{\lambda}|}{2v_F^2n |eB_{\lambda}|+\left(\epsilon_n-s_{B}v_F\lambda p_3\right)^2}} \mathcal{P}_{s_B} e^{ip_3x_3 +ip_2x_2 }\nonumber\\
&\times&\Bigg\{ Y_n(\xi)\left(
                                                        \begin{array}{c}
                                                          0 \\
                                                          1 \\
                                                        \end{array}
                                                      \right)
                                                      - i\frac{\lambda}{v_F\sqrt{2n|eB_{\lambda}|}}\left(\epsilon_{n}-s_{B}v_F\lambda p_3\right)Y_{n-1}(\xi)\left(
                                                        \begin{array}{c}
                                                          1 \\
                                                          0 \\
                                                        \end{array}
                                                      \right)\Bigg\},
\label{model-Bn0-WF-fin-n>0-1}
\end{eqnarray}
\end{subequations}
where
\begin{subequations}
\label{model-Bn0-E-all}
\begin{eqnarray}
\label{model-Bn0-E-be-1}
\epsilon_{n=0} &=& s_{B}v_F\lambda p_3,\\
\epsilon_{n>0} &=&  \pm v_F\sqrt{p_3^2+2n |eB_{\lambda}|}
\label{model-Bn0-E-ee-1}
\end{eqnarray}
\end{subequations}
are the energy dispersion relations at the lowest $n=0$ and higher $n>0$ Landau levels, respectively.
In Eqs.~(\ref{model-Bn0-WF-fin-n=0-1}) and (\ref{model-Bn0-WF-fin-n>0-1}), we used the operator
\begin{equation}
\mathcal{P}_{s_B} = \frac{(1-s_B)}{2} \sigma_x + \frac{(1+s_B)}{2},
\label{model-Bn0-PsB}
\end{equation}
which interchanges spinor components of the wave functions when the sign $s_{B}=\sign{eB_{\lambda}}$ changes. We also used the following shorthand notation:
\begin{equation}
Y_{n}(\xi) = \frac{1}{\sqrt{2^n n! \sqrt{\pi}}}e^{-\xi^2/2}H_{n}(\xi),
\label{model-Bn0-Yn}
\end{equation}
where $H_{n}(\xi)$ denote the Hermite polynomials and $
\xi \equiv \sqrt{|eB_{\lambda}|}\left[x_1 + p_2/(eB_{\lambda})\right]$.

The Wigner function of a many-body system can be defined in terms of the second quantized fermion
and antifermion fields. In a constant magnetic field, the former reads
\begin{equation}
\label{model-Psi-be}
\hat{\Psi} (\mathbf{x}) =  \sum_{n=0}^{\infty} \int \frac{dk_2 dk_3}{(2\pi)^2}\left[
\hat{a}_{n,k_2,k_3} \psi_{n,k_2,k_3}(\mathbf{x})+\hat{b}_{n,k_2,k_3}^{\dag} \phi_{n,k_2,k_3}(\mathbf{x})\right],
\end{equation}
where the summation over the Landau levels and the integration over the corresponding momenta are
performed. Here, $\hat{a}_{n,k_2,k_3}$ denote the particle annihilation operators and
$\hat{b}_{n,k_2,k_3}^{\dag}$ are the antiparticle creation operators. (The Hermitian conjugate
field will be similarly given in terms of particle creation and antiparticle annihilation
operators $\hat{a}_{n,k_2,k_3}^{\dag}$ and $\hat{b}_{n,k_2,k_3}$, respectively.) Both sets of creation
and annihilation operators satisfy the conventional anticommutator relations. While the spinors for
particle states $\psi_{n,k_2,k_3}$ are given by
Eqs.~(\ref{model-Bn0-WF-fin-n=0-1}) and (\ref{model-Bn0-WF-fin-n>0-1}), the spinors for antiparticles
are defined by $\left.\phi_{n,k_2,k_3}\equiv\psi_{n,k_2,k_3}\right|_{\epsilon_n \to -|\epsilon_n|}$.

\section{Wigner function in a constant magnetic field}
\label{sec:Wigner-Weyl-magnetic}

In a relativistic theory, there are several varieties of the Wigner functions. The
Lorentz-covariant forms of the Wigner function were proposed in the context of relativistic quantum
statistical mechanics \cite{Suttorp,deGroot}, as well as in the quantum transport of QCD
\cite{Elze:1986,Vasak:1987,Elze:1989}. However, the covariant formulations lead to
conceptual difficulties when one attempts to solve the kinetic equation as an initial-value problem.
In addition, the physical interpretation of the covariant Wigner function is quite obscure.
The alternative approach is to use the equal-time Wigner function \cite{BialynickiBirula:1991tx}
which breaks explicitly the Lorentz covariance because the Fourier transform with respect to the
relative time coordinate is not performed. However, such an equal-time Wigner function poses a
mathematically well-defined initial-value problem
and its interpretation as a quasiprobability distribution function in the phase space is physically transparent
\cite{BialynickiBirula:1991tx,Gies:2010,Best,Shin}. [Note that the
Wigner function is a quasiprobability distribution function because it can take negative values]. The situation is
quite similar to the study of the Bethe--Salpeter equation which is also fully covariant, but a physical interpretation of the
Bethe--Salpeter two-body wave function is rather obscure.

The equal-time Wigner operator for the Weyl fermions is given by \cite{BialynickiBirula:1991tx}
\begin{equation}
 \hat{W}_{\alpha\eta}(\mathbf{x},\mathbf{p}) = \frac{1}{2}\int d^3 \mathbf{y}\,e^{-i \mathbf{p}\cdot\mathbf{y}} e^{i\Phi(\mathbf{r}_{+},\mathbf{r}_{-})} \left[\hat{\Psi}^{\dag}_{\eta}
 \left(\mathbf{r}_{+}\right), \hat{\Psi}_{\alpha} \left(\mathbf{r}_{-}\right)\right],
\label{Wigner-Weyl-W-def}
\end{equation}
where $\hat{\Psi}_{\alpha}$ and $\hat{\Psi}^{\dag}_{\eta}$ are the spinor components of the chiral fermion fields given
by Eq.~(\ref{model-Psi-be}) and its Hermitian conjugated expression, respectively;
$\mathbf{r}_{\pm}=\mathbf{x}\pm \mathbf{y}/2$, the square brackets denote the commutator, and the
Schwinger phase
\begin{equation}
\Phi(\mathbf{r}_{+},\mathbf{r}_{-})=-e\int_{\mathbf{r}_{-}}^{\mathbf{r}_{+}}d^3\mathbf{r}\mathbf{A}_{\lambda}(\mathbf{r}) = -eB_{\lambda}y_2x_1
\label{Wigner-Weyl-Phi-def}
\end{equation}
ensures the gauge invariance of $\hat{W}_{\alpha\eta}(\mathbf{x},\mathbf{p})$. It is worth noting that instead of the usual
normal ordering, where the vacuum parts are simply omitted, we employed the Schwinger prescription
with a commutator \cite{Schwinger}. In the absence of external fields, both
definitions are physically completely equivalent. However, for time-dependent electromagnetic fields, the definition of normal ordering is ambiguous. In such a case, one should use the Schwinger prescription, which defines the Wigner function correctly transforming
under the charge conjugation.

To obtain a statistical description, the Wigner operator has to be appropriately averaged.
In order to do this, we introduce the density matrix
operator $\hat{\rho}$, which at finite chemical potential and temperature reads
\begin{equation}
\hat{\rho} = \frac{1}{Z} e^{-\beta \left(\hat{H}_{\lambda}-\mu_{\lambda}\hat{N}_{\lambda}\right)},
\label{Wigner-Weyl-rho-def}
\end{equation}
and defines the probability of the realization of a given quantum state. Here, $\beta=1/T$ is the inverse temperature,
$\hat{H}_{\lambda}$ is the second quantized Hamiltonian, $\hat{N}_{\lambda}$ is the
particle number operator, and $Z$ denotes the partition function. By definition,
the latter is given by
\begin{eqnarray}
Z&=& \mbox{Tr}\left[e^{-\beta \left(\hat{H}_{\lambda}-\mu_{\lambda}\hat{N}_{\lambda}\right)}\right] =  \sum_{\Phi} \langle \Phi | e^{-\beta
\left[(|\epsilon_n|-\mu_{\lambda}) \hat{a}^{\dag}_{n, k_2,k_3}\hat{a}_{n,k_2,k_3} -(-|\epsilon_n|
-\mu_{\lambda}) \hat{b}^{\dag}_{n, k_2,k_3}\hat{b}_{n,k_2,k_3}\right]} |\Phi\rangle \nonumber\\
&=& \prod_{n,k_2,k_3} \left(1+e^{-\beta(|\epsilon_n|-\mu_{\lambda})}\right)\left(1+e^{-\beta(|\epsilon_n|
+\mu_{\lambda})}\right),
\label{Wigner-Weyl-Z-def}
\end{eqnarray}
where $\mbox{Tr[\ldots]}$ denotes the trace over the Hilbert space of the multi-particles states
$| \Phi \rangle = | \ldots , N_{m_i}, \ldots , \bar{N}_{m_j}, \ldots \rangle $ with $N_{m_i}$ particles in state $m_i$ and
$\bar{N}_{m_j}$ antiparticles in state $m_j$. Note that the minus sign at the $\hat{b}^{\dag}_{n, k_2,k_3}\hat{b}_{n,k_2,k_3}$ term
in the exponent comes from the normal ordering of the anticommutating fermion operators.

\begin{figure}[tbp]
\centering
\includegraphics[width=.48\textwidth]{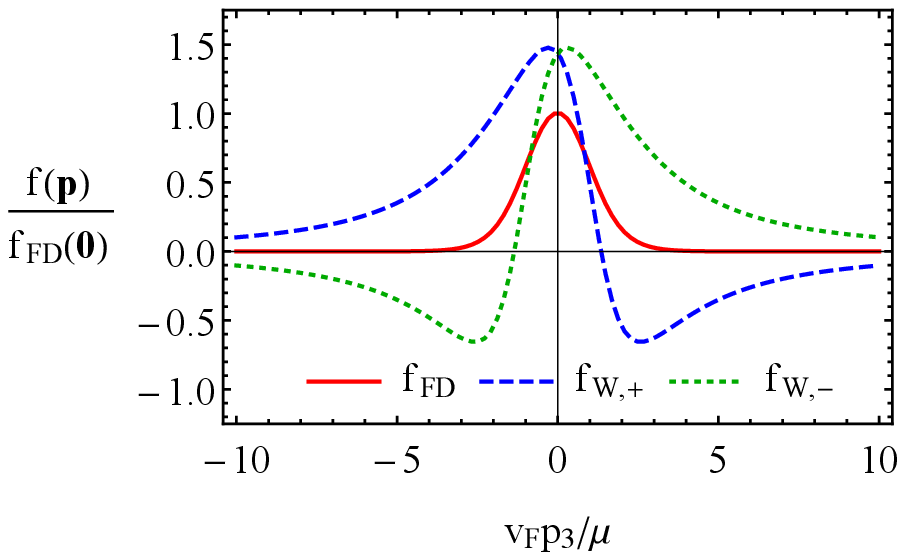}
\hfill
\includegraphics[width=.48\textwidth]{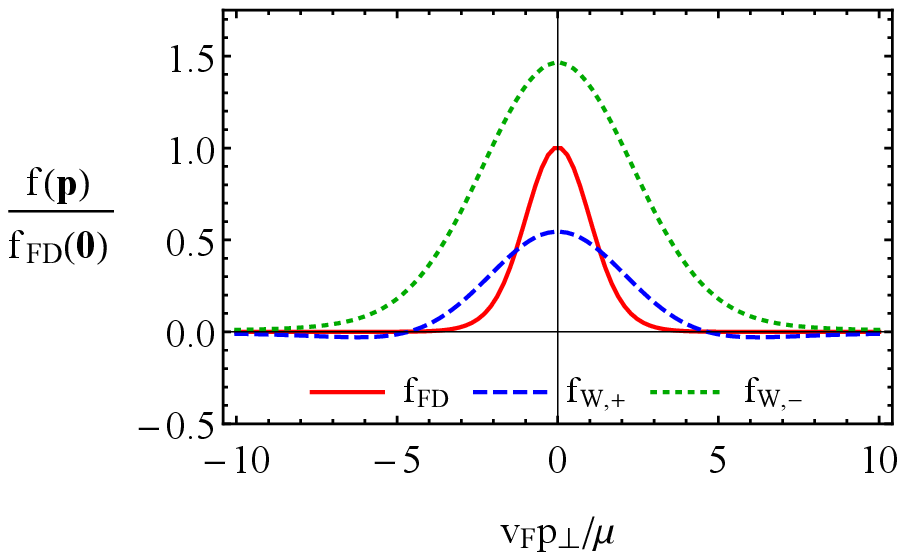}
\caption{\label{fig:Wigner-Weyl-2D}
The dependence of the normalized Wigner quasiprobability distribution functions on $p_3$ (left panel)
and $p_{\perp}$ (right panel) at $B_{\lambda}=0$ (red solid lines) and $B_{\lambda}\neq 0$ (blue dashed
and green doted lines correspond to the right-handed and left-handed fermions, respectively). In the left
panel $p_{\perp}=\mu/v_F$ and $p_{3}=\mu/v_F$ in the right panel. The other parameters are set
as follows: $\mu_5=0$, $T=0.5\,\mu$, $B_{5}=0$, and $eB=10~(\mu/v_F)^2$.}
\end{figure}

By definition, the Wigner function is the ensemble average of the Wigner operator (\ref{Wigner-Weyl-W-def}), i.e.,
\begin{equation}
W_{\alpha\eta}(\mathbf{x},\mathbf{p})\equiv\mbox{Tr}\left(\hat{W}_{\alpha\eta}(\mathbf{x},\mathbf{p}) \hat{\rho}\right).
\label{Wigner-Weyl-W-aver-def}
\end{equation}
Its detailed derivation in the case of chiral fermions in a constant external magnetic field
is presented in appendix~\ref{sec:App-Wigner-der}. The final result can be conveniently
given in the form of an expansion in the Pauli matrices,
\begin{equation}
\label{Wigner-Weyl-w-Pauli-def}
W(\mathbf{x},\mathbf{p})=w_0(\mathbf{x},\mathbf{p})+\lambda \sum_{i=1}^{3}\sigma_i w_i(\mathbf{x},\mathbf{p}).
\end{equation}
Henceforth, we will omit the arguments of $w_0$, $\mathbf{w}$, and $W$. Strictly speaking,
the corresponding functions do not depend on spatial coordinates when the magnetic field is uniform.
The explicit expressions for the vector components of the Wigner function
$\mathbf{w}=\tr{(\bm{\sigma}W)}/2$ are given by
\begin{subequations}
\label{Wigner-Weyl-w-all}
\begin{eqnarray}
\label{Wigner-Weyl-w1}
w_1 &=& 2e^{-p_{\perp}^2/|eB_{\lambda}|} \sum_{n=0}^{\infty} \sum_{\epsilon_n}(-1)^n \frac{v_F p_1}{|\epsilon_n|}L_{n-1}^{1}\left(\frac{2p_{\perp}^2}{|eB_{\lambda}|}\right)
\Bigg\{\frac{\theta(\epsilon_n)}{1+e^{\beta(|\epsilon_n|-\mu_{\lambda})}}
+\frac{\theta(-\epsilon_n)}{1+e^{\beta(|\epsilon_n|+\mu_{\lambda})}} -\frac{1}{2}\Bigg\},
\nonumber\\
&&\\
\label{Wigner-Weyl-w2}
w_2 &=& 2e^{-p_{\perp}^2/|eB_{\lambda}|} \sum_{n=0}^{\infty} \sum_{\epsilon_n}(-1)^n \frac{v_F p_2}{|\epsilon_n|}L_{n-1}^{1}\left(\frac{2p_{\perp}^2}{|eB_{\lambda}|}\right)
\Bigg\{\frac{\theta(\epsilon_n)}{1+e^{\beta(|\epsilon_n|-\mu_{\lambda})}}
+\frac{\theta(-\epsilon_n)}{1+e^{\beta(|\epsilon_n|+\mu_{\lambda})}} -\frac{1}{2}\Bigg\},\nonumber\\
&&\\
\label{Wigner-Weyl-w3}
w_3 &=& \lambda s_B\frac{e^{-p_{\perp}^2/|eB_{\lambda}|}}{2} \sum_{n=0}^{\infty} \sum_{\epsilon_n} \frac{(-1)^n}{|\epsilon_n|}\nonumber\\
&\times&
\Bigg\{-\Bigg[(|\epsilon_n|+s_B\lambda v_F p_3)L_{n}\left(\frac{2p_{\perp}^2}{|eB_{\lambda}|}\right)
+(|\epsilon_n|-s_B\lambda v_F p_3)L_{n-1}\left(\frac{2p_{\perp}^2}{|eB_{\lambda}|}\right)\Bigg] \frac{\theta(\epsilon_n)}{1+e^{\beta(|\epsilon_n|-\mu_{\lambda})}} \nonumber\\
&+&\left[(|\epsilon_n|-s_B\lambda v_F p_3)L_{n}\left(\frac{2p_{\perp}^2}{|eB_{\lambda}|}\right) +(|\epsilon_n|+s_B\lambda v_F p_3)L_{n-1}\left(\frac{2p_{\perp}^2}{|eB_{\lambda}|}\right)\right] \frac{\theta(-\epsilon_n)}{1+e^{\beta(|\epsilon_n|+\mu_{\lambda})}} \nonumber\\
&+&\frac{1}{2}\left[(\epsilon_n+s_B\lambda v_F p_3)L_{n}\left(\frac{2p_{\perp}^2}{|eB_{\lambda}|}\right) +(\epsilon_n-s_B\lambda v_F p_3)L_{n-1}\left(\frac{2p_{\perp}^2}{|eB_{\lambda}|}\right)\right]  \Bigg\},
\end{eqnarray}
\end{subequations}
where $p_{\perp}^2=p_1^2+p_2^2$, $\theta(x)$ denotes the unit step function, and the sum
$\sum_{\epsilon_n}$ takes into account both positive and negative branches of the
energy spectrum at $n>0$.
The scalar part of the Wigner function $w_0=\tr{\left(W\right)}/2$ defines the quasiprobability distribution function
$f_{\rm W,\lambda}(\mathbf{p})\equiv2w_0$. The explicit expression of the latter reads
\begin{eqnarray}
\label{Wigner-Weyl-B-W-mean}
f_{\rm W,\lambda}(\mathbf{p})&=& 2 e^{-p_{\perp}^2/|eB_{\lambda}|} \Bigg\{\frac{\theta(\epsilon_0)}{1+e^{\beta(|\epsilon_0|-\mu_{\lambda})}}-\frac{\theta(-\epsilon_0)}{1+e^{\beta(|\epsilon_0|+\mu_{\lambda})}} -\frac{\sign{\epsilon_0}}{2} \Bigg\}
+ e^{-p_{\perp}^2/|eB_{\lambda}|} \sum_{n=1}^{\infty} \frac{(-1)^n}{|\epsilon_n|}\nonumber\\
&\times&
\Bigg\{\Bigg[(|\epsilon_n|+s_B\lambda v_F p_3)L_{n}\left(\frac{2p_{\perp}^2}{|eB_{\lambda}|}\right)
-(|\epsilon_n|-s_B\lambda v_F p_3)L_{n-1}\left(\frac{2p_{\perp}^2}{|eB_{\lambda}|}\right)\Bigg] \frac{1}{1+e^{\beta(|\epsilon_n|-\mu_{\lambda})}} \nonumber\\
&-&\left[(|\epsilon_n|-s_B\lambda v_F p_3)L_{n}\left(\frac{2p_{\perp}^2}{|eB_{\lambda}|}\right) -(|\epsilon_n|+s_B\lambda v_F p_3)L_{n-1}\left(\frac{2p_{\perp}^2}{|eB_{\lambda}|}\right)\right] \frac{1}{1+e^{\beta(|\epsilon_n|+\mu_{\lambda})}} \nonumber\\
&-&s_B\lambda v_F p_3\left[L_{n}\left(\frac{2p_{\perp}^2}{|eB_{\lambda}|}\right) +L_{n-1}\left(\frac{2p_{\perp}^2}{|eB_{\lambda}|}\right)\right]
\Bigg\}.
\end{eqnarray}
In order to get a better insight into the Wigner quasiprobability distribution function
$f_{\rm W, \lambda}(\mathbf{p})$, it is instructive to compare it with the standard
quasiprobability $f_{\rm FD}(\mathbf{p})$ at $B_{\lambda}=0$, which is given in terms of the
Fermi-Dirac functions, i.e.,
\begin{equation}
\label{Wigner-Weyl-B-f-FD}
f_{\rm FD}(\mathbf{p})= n_{\rm F}(\epsilon_{\mathbf{p}}-\mu_{\lambda})-n_{\rm F}(\epsilon_{\mathbf{p}}+\mu_{\lambda})
\end{equation}
where $\epsilon_{\mathbf{p}}=v_F|\mathbf{p}|$ and $n_{\rm F}(x)=1/(e^{\beta x}+1)$ is the
Fermi-Dirac distribution function.
The numerical comparison of the two
quasiprobability distribution functions is shown in Fig.~\ref{fig:Wigner-Weyl-2D}. As we can see,
the inclusion of the magnetic field leads to several qualitative changes in the dependence
of the quasiprobabilities on the longitudinal and transverse momenta, presented in the left and
right panels of figure~\ref{fig:Wigner-Weyl-2D}, respectively. While the quasiprobability function
(\ref{Wigner-Weyl-B-f-FD}) is always positive (assuming $\mu_\lambda>0$), its
counterpart in the background magnetic field takes negative values in a range of momenta.
Such negative values of the quasiprobability originate from the quantum effects that cannot be captured by usual distribution functions.
As is seen from the left panel in figure~\ref{fig:Wigner-Weyl-2D}, the dependence of the quasiprobability
distribution function $f_{\rm W, \lambda}(\mathbf{p})$ on $p_3$ is asymmetric in the longitudinal component of
momentum $p_3$, as well as in chirality. A chiral asymmetry is also clearly visible in the right panel in
figure~\ref{fig:Wigner-Weyl-2D}, where the distributions $f_{\rm W, \pm}(\mathbf{p})$ have different widths
and heights as functions of $p_\perp$.

\begin{figure}[tbp]
\centering
\includegraphics[width=.48\textwidth]{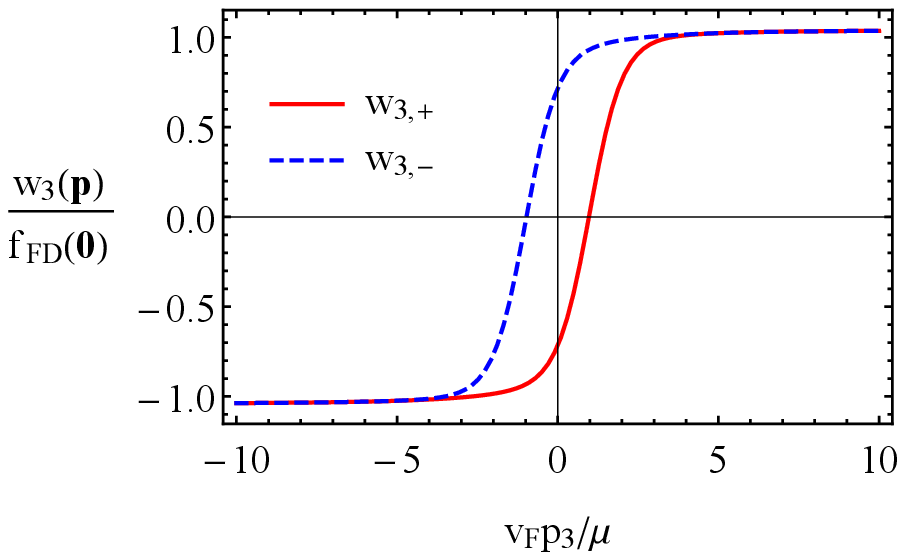}
\hfill
\includegraphics[width=.48\textwidth]{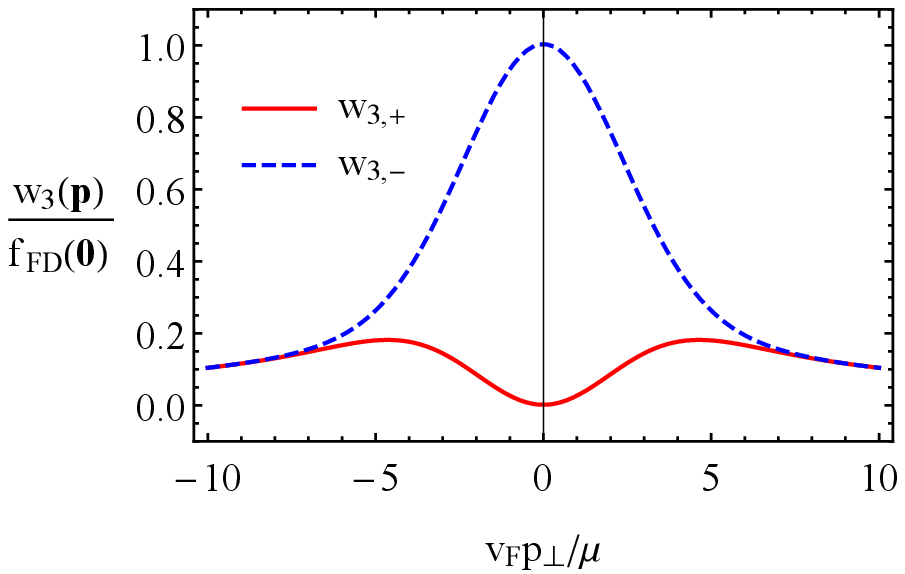}
\caption{\label{fig:Wigner-Weyl-w3}
The dependence of the normalized vector component of the Wigner function
along the magnetic field defined by Eq.~(\ref{Wigner-Weyl-w3}) on $p_3$ (left panel) and
$p_{\perp}$ (right panel). Red solid and blue dashed lines correspond to the right-handed and
left-handed fermions, respectively. In the left panel $p_{\perp}=\mu/v_F$ and $p_{3}=\mu/v_F$
in the right panel. The other parameters are set as follows: $\mu_5=0$, $T=0.5\,\mu$, $B_{5}=0$, and
$eB=10~(\mu/v_F)^2$.}
\end{figure}

Further, we plot the dependence of the vector component of the Wigner function along the magnetic field $w_3$ on $p_3$ and $p_{\perp}$ in the left and right panels of figure~\ref{fig:Wigner-Weyl-w3}, respectively. Similarly to the Wigner quasiprobability distribution function, the corresponding dependence is also asymmetric with respect to $p_3$ and the chirality. Note that the asymmetry is well-pronounced at small values of momenta.

Before proceeding further with the analysis of the Wigner function in a strong magnetic field, we will test
our results (\ref{Wigner-Weyl-w-Pauli-def})--(\ref{Wigner-Weyl-B-W-mean}) by studying the chiral magnetic effect (CME) and chiral separation effect (CSE), as well as
deriving the weak-field limit of the function $W$.

\subsection{Chiral magnetic and chiral separation effects}
\label{sec:Wigner-Weyl-CME-CSE}

We begin our analysis of the Wigner function (\ref{Wigner-Weyl-w-Pauli-def}) with the study of the chiral magnetic and chiral
separation effects. The electric and chiral current densities are defined by
\begin{subequations}
\label{Wigner-Weyl-CME-CSE-j-j5-def}
\begin{eqnarray}
\label{Wigner-Weyl-CME-CSE-j-def}
\mathbf{j} &\equiv& -ev_F\sum_{\lambda=\pm} \lambda\int \frac{d^3\mathbf{p}}{(2\pi)^3} \mbox{tr}\left[\bm{\sigma}W\right]
= -2ev_F\sum_{\lambda=\pm} \int \frac{d^3\mathbf{p}}{(2\pi)^3} \mathbf{w},\\
\label{Wigner-Weyl-CME-CSE-j5-def}
\mathbf{j}_5 &\equiv& -ev_F\sum_{\lambda=\pm} \int \frac{d^3\mathbf{p}}{(2\pi)^3} \mbox{tr}\left[\bm{\sigma}W\right]
= -2ev_F\sum_{\lambda=\pm} \lambda\int \frac{d^3\mathbf{p}}{(2\pi)^3} \mathbf{w}.
\end{eqnarray}
\end{subequations}
It is worth noting that the factor $\lambda$ in Eq.~(\ref{Wigner-Weyl-CME-CSE-j-def}) comes from the definition of the electric
current operator
\begin{equation}
j_{i,\lambda}(\mathbf{p})\equiv-e(\partial_{p_i}H_{\lambda})=-e\lambda v_F\sigma_i.
\label{Wigner-Weyl-CME-CSE-j-op-def}
\end{equation}
Taking into account the Wigner function components given by Eqs.~(\ref{Wigner-Weyl-w1})--(\ref{Wigner-Weyl-w3}) one can easily see that the only
nonzero component of the current is along the $\hat{\mathbf{z}}$ direction. Integrating over $p_{\perp}$, we obtain
\begin{eqnarray}
\label{Wigner-Weyl-CME-CSE-j-int-W}
j_{3,\lambda}&=& \lambda ev_F\frac{eB_{\lambda}}{(2\pi)^2} \int dp_3 \left[\frac{\theta(s_B\lambda v_Fp_3)}{1+e^{\beta(v_F|p_3|-\mu_{\lambda})}}-\frac{\theta(-s_B\lambda v_Fp_3)}{1+e^{\beta(v_F|p_3|+\mu_{\lambda})}} -s_B\lambda \frac{\sign{p_3}}{2}\right]  \nonumber\\
&+& ev_F\frac{|eB_{\lambda}|}{(2\pi)^2} \sum_{n=1}^{\infty} \int dp_3
\frac{p_3}{\sqrt{p_3^2+2n|eB_{\lambda}|}}
\left[  \frac{\theta(\epsilon_n)}{1+e^{\beta(|\epsilon_n|-\mu_{\lambda})}} + \frac{\theta(-\epsilon_n)}{1+e^{\beta(|\epsilon_n|+\mu_{\lambda})}} - \frac{1}{2} \right]
 \nonumber\\
&=& \lambda e\frac{eB_{\lambda} }{(2\pi)^2} \int dp_3 T\left[\ln{\left(1+e^{\beta \mu_{\lambda}}\right)} -\ln{\left(1+e^{-\beta \mu_{\lambda}}\right)}\right] =  \lambda \frac{e^2B_{\lambda} \mu_{\lambda} }{(2\pi)^2},
\end{eqnarray}
where the contribution from the higher Landau levels is zero due to the integration over $p_3$. Performing the summation over chiralities, we find the following standard
electric and chiral current densities \cite{CME,Vilenkin:1980ft,Zhitnitsky,Grushin-Vishwanath:2016}:
\begin{eqnarray}
\label{Wigner-Weyl-CME-CSE-answer-be}
j_{3} &=& \sum_{\lambda}j_{3,\lambda} = \frac{e^2B \mu_5}{2\pi^2}+\frac{e^2B_5 \mu}{2\pi^2},\\
\label{Wigner-Weyl-CME-CSE-answer-ee}
j_{3,5} &=& \sum_{\lambda}\lambda j_{3,\lambda} =  \frac{e^2B \mu}{2\pi^2} +\frac{e^2B_5 \mu_5}{2\pi^2}.
\end{eqnarray}
Thus, as expected, we reproduce exactly the standard relations for the CSE and CME using the Wigner
function approach.

\subsection{Weak magnetic field expansion}
\label{sec:Wigner-Weyl-weak-B-III}

In this subsection we consider the limit of small magnetic fields $|eB_{\lambda}|\ll p^2$. After expanding
the Wigner function to the linear order in $|eB_{\lambda}|/p^2$ and performing the summation over the
Landau levels (see appendix~\ref{sec:App-Wigner-weak-B} for details), we arrive at the following expressions
for the scalar $w_0$ and vector $\mathbf{w}$ parts:
\begin{subequations}
\label{Wigner-C-wB-w-all}
\begin{eqnarray}
\label{Wigner-C-wB-w-be}
w_0&=&\frac{1}{2} \left[n_{\rm F}(\epsilon_{\mathbf{p}}-\mu_{\lambda}) -n_{\rm F}(\epsilon_{\mathbf{p}}+\mu_{\lambda})\right]\nonumber \\
&-&\frac{\lambda v_F^2eB_{\lambda}}{4} \frac{v_Fp_3}{\epsilon_{\mathbf{p}}} \frac{d}{d\epsilon_{\mathbf{p}}} \frac{1}{\epsilon_{\mathbf{p}}}\Big[n_{\rm F}(\epsilon_{\mathbf{p}}-\mu_{\lambda})+n_{\rm F}(\epsilon_{\mathbf{p}}+\mu_{\lambda}) - 1\Big]  +O\left(|eB_{\lambda}|^2\right),\\
w_j&=& -\frac{v_F p_j}{2} \frac{1}{\epsilon_{\mathbf{p}}} \left[n_{\rm F}(\epsilon_{\mathbf{p}}-\mu_{\lambda})+n_{\rm F}(\epsilon_{\mathbf{p}}+\mu_{\lambda}) -1\right]\nonumber \\
&+& \frac{\lambda v_F^2e(B_{\lambda})_j}{4} \frac{1}{\epsilon_{\mathbf{p}}} \frac{d}{d\epsilon_{\mathbf{p}}} \Big[n_{\rm F}(\epsilon_{\mathbf{p}}-\mu_{\lambda})-n_{\rm F}(\epsilon_{\mathbf{p}}+\mu_{\lambda})\Big]
+O\left(|eB_{\lambda}|^2\right).
\label{Wigner-C-wB-w-ee}
\end{eqnarray}
\end{subequations}
The above expressions qualitatively agree with the results obtained in Refs.~\cite{Gao:2012ix,Wang-kinetic,Gao:2017rgi}. Note, however, that those papers use the covariant definition of the Wigner function.
Therefore, one needs to integrate their results over the zeroth component of the four-momentum $p_0$
before comparing with Eqs.~(\ref{Wigner-C-wB-w-be}) and (\ref{Wigner-C-wB-w-ee}).
It is interesting to note that we have the additional term $-1$ in the second square brackets
of Eq.~(\ref{Wigner-C-wB-w-be}) as well as in the first square brackets of Eq.~(\ref{Wigner-C-wB-w-ee}),
which is absent in Refs.~\cite{Gao:2012ix,Wang-kinetic,Gao:2017rgi}. This difference is connected
with our use of the commutator in the definition of the Wigner operator (\ref{Wigner-Weyl-W-def})
instead of the usual normal ordering considered in the cited works.

By making use of the explicit expressions for the Fermi-Dirac distribution functions,
the results for the scalar and vector parts of the Wigner function can be rewritten in the following
equivalent form:
\begin{eqnarray}
\label{Wigner-Weyl-compar-W-CKT}
w_0 &\simeq& \frac{1}{2}  \Bigg\{\left[1+e(\mathbf{B}_{\lambda}\cdot\mathbf{\Omega}_{\lambda})\right] n_{\rm F}\left(\epsilon_{\mathbf{p}}-\frac{\lambda ev_F (\mathbf{B}_{\lambda}\cdot\mathbf{p})}{2|\mathbf{p}|^2}-\mu_{\lambda}\right) \nonumber\\
&-&\left[1-e(\mathbf{B}_{\lambda}\cdot\mathbf{\Omega}_{\lambda})\right] n_{\rm F}\left(\epsilon_{\mathbf{p}}+\frac{\lambda ev_F (\mathbf{B}_{\lambda}\cdot\mathbf{p})}{2|\mathbf{p}|^2}+\mu_{\lambda}\right)\Bigg\}
- \frac{e(\mathbf{B}_{\lambda}\cdot\mathbf{\Omega}_{\lambda})}{2}  +O\left(|eB_{\lambda}|^2\right),\nonumber\\
\end{eqnarray}
where $\mathbf{\Omega}_{\lambda} = \lambda \mathbf{p}/(2|\mathbf{p}|^3)$ is the Berry curvature.
This result agrees
with the distribution function in the chiral kinetic theory, except for the last term, which originates from the commutator in the definition of the Wigner operator (\ref{Wigner-Weyl-W-def}). Note that due to the presence of the Berry curvature and magnetic field, the quasiprobability distribution function in Eq.~(\ref{Wigner-Weyl-compar-W-CKT}) can take negative values when the magnetic field is nonzero.

\section{Equation of motion for the Wigner function}
\label{sec:Wigner-Weyl-Eq}

In this section we present the equation of motion for the equal-time Wigner function in external electromagnetic
fields. According to Ref.~\cite{BialynickiBirula:1991tx}, the corresponding equation reads
\begin{equation}
\label{Wigner-Weyl-Eq-def}
D_0W+v_F\frac{\lambda}{2}\mathbf{D}\cdot\left\{\bm{\sigma},W\right\}-i\lambda v_F \left[(\bm{\sigma}\cdot\mathbf{P}), W\right]=0.
\end{equation}
Here $\left\{,\right\}$ and $\left[,\right]$ denote anticommutator and commutator, respectively, and the following derivatives are used:
\begin{subequations}
\label{Wigner-Weyl-Eq-D-all}
\begin{eqnarray}
\label{Wigner-Weyl-Eq-D-be}
D_0 &\equiv&  \partial_t+e\int_{-1/2}^{1/2}ds \,\left(\mathbf{E}(\mathbf{r}+is\partial_{\mathbf{p}})\cdot\partial_{\mathbf{p}} \right) \approx\partial_t+e\left(\mathbf{E}(\mathbf{r})\cdot\partial_{\mathbf{p}}\right) +O\left[(\nabla_{\mathbf{r}}\cdot\partial_{\mathbf{p}})\mathbf{E}(\mathbf{r})\right],\\
\mathbf{D} &\equiv& \partial_\mathbf{r}+e\int_{-1/2}^{1/2}ds \, \left[\mathbf{B}(\mathbf{r}+is\partial_{\mathbf{p}})\times\partial_{\mathbf{p}}\right]\approx\partial_\mathbf{r}+e[\mathbf{B}(\mathbf{r})\times\partial_{\mathbf{p}}] +O\left[(\nabla_{\mathbf{r}}\times\partial_{\mathbf{p}})\mathbf{B}(\mathbf{r})\right],\nonumber\\
\\
\mathbf{P} &\equiv&\mathbf{p} +ie\int_{-1/2}^{1/2}ds\,s\left[\mathbf{B}(\mathbf{r}+is\partial_{\mathbf{p}})\times\partial_{\mathbf{p}}\right] \approx \mathbf{p} +O\left[(\nabla_{\mathbf{r}}\cdot\partial_{\mathbf{p}})\mathbf{B}(\mathbf{r})\right].
\label{Wigner-Weyl-Eq-D-ee}
\end{eqnarray}
\end{subequations}
As one can see from the above equations, the derivatives become local when the external fields are spatially uniform. In terms of
the scalar $w_0$ and vector $\mathbf{w}$ parts of the Wigner function, Eq.~(\ref{Wigner-Weyl-Eq-def})
reads
\begin{subequations}
\label{Wigner-Weyl-Eq-eqs-all}
\begin{eqnarray}
\label{Wigner-Weyl-Eq-eqs-be}
D_0 w_0+v_F\left(\mathbf{D}\cdot\mathbf{w}\right)=0,\\
D_0\mathbf{w}+v_F\mathbf{D} w_0+2\lambda v_F[\mathbf{P}\times \mathbf{w}]=0.
\label{Wigner-Weyl-Eq-eqs-ee}
\end{eqnarray}
\end{subequations}
In the next section, we will use Eqs.~(\ref{Wigner-Weyl-Eq-eqs-be}) and (\ref{Wigner-Weyl-Eq-eqs-ee})
to study the longitudinal modes of the chiral magnetic wave (CMW) \cite{Yee} in the limit of
a strong magnetic field. In order to describe such a collective excitation taking into account the dynamical electromagnetism, we consider the system subjected to small oscillating electromagnetic fields
\begin{subequations}
\label{Wigner-Weyl-Eq-EB-prime}
\begin{eqnarray}
\mathbf{E}^{\prime} &=& \mathbf{E} e^{-i\omega t +i\mathbf{k}\cdot\mathbf{r}}, \\
\mathbf{B}^{\prime} &=& \mathbf{B}
e^{-i\omega t +i\mathbf{k}\cdot\mathbf{r}},
\end{eqnarray}
\end{subequations}
and a strong constant effective magnetic field $\mathbf{B}_{0,\lambda}$. [In the case of a weak magnetic field,
the effects of the dynamical electromagnetism were taken into account in Refs.~\cite{Gorbar:2016ygi,Gorbar:2016sey}.]
In this case, the Wigner function can be naturally split in two parts, i.e., $W = W^{(0)} + W^{\prime}$.
While the first part corresponds to the constant external magnetic field $\mathbf{B}_{0,\lambda}$, the second one is
related to the oscillating fields. The latter can be written in the form
\begin{equation}
\label{Wigner-Weyl-Eq-W-prime}
W^{\prime}=W^{(1)}e^{-i\omega t +i\mathbf{k}\cdot\mathbf{r}}=w_0^{\prime}+\lambda(\bm{\sigma}\cdot\mathbf{w}^{\prime})= w_0^{(1)}e^{-i\omega t +i\mathbf{k}\cdot\mathbf{r}} +\lambda(\bm{\sigma}\cdot\mathbf{w}^{(1)})e^{-i\omega t +i\mathbf{k}\cdot\mathbf{r}}.
\end{equation}
To the linear order in the oscillating electromagnetic fields, the components of the Wigner function satisfy the
following equations:
\begin{subequations}
\label{Wigner-Weyl-Eq-eqs-oscil-all}
\begin{eqnarray}
\label{Wigner-Weyl-Eq-eqs-oscil-be}
&&\partial_{t}w_0^{\prime} +v_F\partial_{\mathbf{r}}\mathbf{w}^{\prime} +v_Fe\left(\left[\mathbf{B}_{0,\lambda}\times\partial_{\mathbf{p}}\right]\cdot\mathbf{w}^{\prime}\right) + e(\mathbf{E}^{\prime}\cdot\partial_{\mathbf{p}})\int_{-1/2}^{1/2}ds\, w_0^{(0)}(\mathbf{p}-s\mathbf{k}) \nonumber\\
&&+v_Fe\left[\mathbf{B}^{\prime}\times\partial_{\mathbf{p}}\right]\int_{-1/2}^{1/2}ds\, \mathbf{w}^{(0)}(\mathbf{p}-s\mathbf{k}) =0,\\
&&\partial_{t}\mathbf{w}^{\prime}+v_F\partial_{\mathbf{r}}w_0^{\prime}  +v_Fe\left[\mathbf{B}_{0,\lambda}\times\partial_{\mathbf{p}}\right]w_0^{\prime} + 2\lambda v_F\left[\mathbf{p}\times\mathbf{w}^{\prime}\right] \nonumber\\
&&-v_Fe\left[\mathbf{B}^{\prime}\times\partial_{\mathbf{p}}\right]\int_{-1/2}^{1/2}ds\, w_0^{(0)}(\mathbf{p}-s\mathbf{k})
+ e(\mathbf{E}^{\prime}\cdot\partial_{\mathbf{p}})\int_{-1/2}^{1/2}ds \mathbf{w}^{(0)}(\mathbf{p}-s\mathbf{k}) \nonumber\\
&&+2ie\lambda v_F \int_{-1/2}^{1/2}ds\,s\left[\left[\mathbf{B}^{\prime} \times\partial_{\mathbf{p}}\right]\times\mathbf{w}^{(0)}(\mathbf{p}-s\mathbf{k})\right]  =0.
\label{Wigner-Weyl-Eq-eqs-oscil-ee}
\end{eqnarray}
\end{subequations}
In order to find the spectrum of collective modes, we should determine the electric current density $j^{\prime m}$ which enters the Maxwell's equations
through the polarization vector
\begin{equation}
P^{\prime m}= i\frac{j^{\prime m}}{\omega}=\chi^{mn}E^{\prime n},
\label{second-polarization-tensor}
\end{equation}
where $\chi^{mn}$ ($m,n=1,2,3$ denote spatial components) is the susceptibility tensor.
Then, as is easy to check, the Maxwell's equations admit a nontrivial solution when the following characteristic
equation is satisfied:
\begin{equation}
\mbox{det}\left[\left(\omega^2- k^2 \right)\delta^{mn} + k^m k^n + 4\pi\omega^2\chi^{mn}\right]=0.
\label{second-collective-B-tensor-dispersion-relation-general}
\end{equation}
The solution to this equation determines the dispersion relation of electromagnetic collective
modes, such as the chiral magnetic wave.

As we saw in the previous section, the scalar part of the Wigner function is related to the distribution
function of the chiral kinetic theory in the limit of weak magnetic field. However, the Wigner
function is applicable even beyond the weak-field limit. It is instructive, therefore, to consider the case
of a strong constant magnetic field $\mathbf{B}_{0,\lambda}\parallel \hat{\mathbf{z}}$. In such a limit,
one can use the lowest Landau level (LLL) approximation when only the LLL contribution is retained.
Then, the scalar and vector components of the Wigner function take the form
\begin{subequations}
\label{Wigner-C-LLL-w-all}
\begin{eqnarray}
\label{Wigner-C-LLL-w0}
w_{0}^{(0)}&=&  e^{-p_{\perp}^2/|eB_{0,\lambda}|} \tilde{f}_{\rm LLL}(p_3),\\
\mathbf{w}^{(0)} &=& -s_{B}\lambda\hat{\mathbf{z}} e^{-p_{\perp}^2/|eB_{0,\lambda}|} \tilde{f}_{\rm LLL}(p_3),
\label{Wigner-C-LLL-wvec}
\end{eqnarray}
\end{subequations}
where
\begin{equation}
\label{Wigner-C-LLL-fLLL}
\tilde{f}_{\rm LLL}(p_3) \equiv \theta(s_B\lambda v_Fp_3)n_{\rm F} (s_B\lambda v_Fp_3-\mu_{\lambda}) - \theta(-s_B\lambda v_Fp_3)
n_{\rm F} (-s_B\lambda v_Fp_3+\mu_{\lambda})
-\frac{\lambda s_B\sign{p_3}}{2}.
\end{equation}
Note that both scalar and vector parts of the Wigner function are expressed in terms of the distribution
function on the LLL. The last term in Eq.~(\ref{Wigner-C-LLL-fLLL}) is related to our use of the commutator in
the Wigner operator (\ref{Wigner-Weyl-W-def}) and properly describes the vacuum oscillations. As we will see
below, it is crucial for the correct description of the collective excitations and transport phenomena in a strong
field limit.

In the next two sections, we will use the equal-time Wigner function in the LLL approximation
in order to study: (i) the dispersion relation of the CMW and (ii) the thermoelectric properties of chiral fermions
in a strong magnetic field.

\section{The chiral magnetic and pseudomagnetic waves in a strong magnetic field}
\label{sec:Wigner-Weyl-LLL-plasmon}

In this section, we study the dispersion relation of the CMW in the strong magnetic field limit by using the
equal-time Wigner function approach. In order to simplify the analysis, we will consider only the case of
longitudinal waves, propagating along the direction of the background field. In other words, the perpendicular
components of the wave vector and the oscillating electric field will vanish, i.e., $\mathbf{k}_{\perp}=0$
and $\mathbf{E}_{\perp}^{\prime}=0$. In view of the Maxwell equations, there will be also no oscillating
magnetic field, i.e., $\mathbf{B}^{\prime}=[\mathbf{k}\times \mathbf{E}^{\prime}]/\omega=0$. In this case,
Eqs.~(\ref{Wigner-Weyl-Eq-eqs-oscil-be}) and (\ref{Wigner-Weyl-Eq-eqs-oscil-ee}) reduce to
\begin{subequations}
\label{Wigner-C-eqs-1-S-all}
\begin{eqnarray}
\label{Wigner-C-eqs-1-S-be}
&&-i\omega w_0^{(1)} +iv_Fk_3w_3^{(1)}  + eE_3 e^{-p_{\perp}^2/|eB_{0,\lambda}|} \int_{-1/2}^{1/2}ds \partial_{p_3}\tilde{f}_{\rm LLL}(p_3-sk_3) =0,\\
&&-i\omega w_3^{(1)}  +iv_Fk_3 w_0^{(1)}  -s_B\lambda eE_3 e^{-p_{\perp}^2/|eB_{0,\lambda}|} \int_{-1/2}^{1/2}ds \partial_{p_3}\tilde{f}_{\rm LLL}(p_3-sk_3)=0.
\label{Wigner-C-eqs-1-S-ee}
\end{eqnarray}
\end{subequations}
Note that due to the one-dimensional nature of the LLL, the equations for the $w_1^{(1)}$ and $w_2^{(1)}$
have only the trivial solutions $w_1^{(1)} = w_2^{(1)} =0$. On the other hand, the solution to the system of coupled equations (\ref{Wigner-C-eqs-1-S-be}) and (\ref{Wigner-C-eqs-1-S-ee}) is nontrivial, i.e.,
\begin{subequations}
\label{Wigner-C-sol-S-all}
\begin{eqnarray}
\label{Wigner-C-sol-S-be}
w_0^{(1)} &=& -ieE_3 e^{-p_{\perp}^2/|eB_{0,\lambda}|} \int_{-1/2}^{1/2}ds  \frac{\partial_{p_3}\tilde{f}_{\rm LLL}(p_3-sk_3)}{\omega+\lambda s_Bv_Fk_3},\\
w_3^{(1)} &=& s_B\lambda ieE_3 e^{-p_{\perp}^2/|eB_{0,\lambda}|} \int_{-1/2}^{1/2}ds  \frac{\partial_{p_3}\tilde{f}_{\rm LLL}(p_3-sk_3)}{\omega+\lambda s_Bv_Fk_3}.
\label{Wigner-C-sol-S-ee}
\end{eqnarray}
\end{subequations}
By making use of the definition in Eq.~(\ref{Wigner-Weyl-CME-CSE-j-def}), we then derive the following
correction to the electric current density proportional to the oscillating electric field:
\begin{equation}
j_{3}^\prime = -2i e^2v_F  \sum_{\lambda=\pm}s_B\lambda E_3^\prime \int \frac{d^3\mathbf{p}}{(2\pi)^3}  \frac{e^{-p_{\perp}^2/|eB_{0,\lambda}|}}{\omega+\lambda s_Bv_Fk_3} \int_{-1/2}^{1/2}ds \partial_{p_3}\tilde{f}_{\rm LLL}(p_3-sk_3).
\label{Wigner-C-j3-0}
\end{equation}
It should be noted that there is also a non-oscillating contribution to the current that comes from
the zeroth order Wigner function $w_3^{(0)}$. It describes the chiral magnetic and chiral separation effects,
but does not affect directly the dispersion relation of the collective modes.

In order to perform the integral over $s$ on the right-hand side of Eq.~(\ref{Wigner-C-j3-0}) it is
convenient to rewrite the partial derivative with respect to $p_3$ in terms of the partial derivative
with respect to $s$, i.e.,
\begin{eqnarray}
\label{Wigner-C-EB-trick}
I_{k_3}&\equiv& \int_{-\Lambda}^{\Lambda}dp_3\int_{-1/2}^{1/2}ds \partial_{p_3}\tilde{f}_{\rm LLL}(p_3-sk_3) = \int_{-\Lambda}^{\Lambda}dp_3\int_{-1/2}^{1/2}ds \frac{1}{-k_3}\partial_s\tilde{f}_{\rm LLL}(p_3-sk_3) \nonumber\\
&=&2s_B\lambda +\frac{2T}{s_B\lambda v_Fk_3}\Bigg\{\theta(s_B\lambda) \frac{-s_B\lambda v_Fk_3}{T} +\theta(-s_B\lambda) \frac{s_B\lambda v_Fk_3}{T}\Bigg\} -s_B\lambda \nonumber\\
&=&-s_B\lambda.
\end{eqnarray}
Then, the final result for the electric current density (\ref{Wigner-C-j3-0}) reads
\begin{eqnarray}
j_{3}^\prime  &=& -2ie^2v_F \sum_{\lambda=\pm}\lambda s_B e E_3^\prime \int \frac{d^2\mathbf{p}_{\perp} e^{-p_{\perp}^2/|eB_{0,\lambda}|} }{(2\pi)^3}  \frac{I_{k_3}}{\omega+\lambda s_Bv_Fk_3} \nonumber\\
&=&ie^2v_F \sum_{\lambda=\pm} \frac{E_3^\prime}{\omega+\lambda s_Bv_Fk_3} \frac{|eB_{0,\lambda}|}{(2\pi)^2}.
\label{Wigner-C-j3-1}
\end{eqnarray}

Let us first consider the case of an ordinary magnetic field background, i.e., $B_{0}\neq 0$ but
$B_{0,5}=0$. By comparing with Eq.~(\ref{second-polarization-tensor}), we extract the following susceptibility
tensor:
\begin{equation}
\label{Wigner-C-P3}
\chi^{33} = -\frac{v_Fe^2|eB_0|}{2\pi^2(\omega^2-v_F^2k_3^2)}.
\end{equation}
By substituting this into the characteristic equation
(\ref{second-collective-B-tensor-dispersion-relation-general}),
we then derive the following positive-energy solution for collective modes:
\begin{equation}
\label{Wigner-C-omega-B-sol}
\omega=\sqrt{v_F^2k_3^2+\frac{2v_Fe^2|eB_{0}|}{\pi}}.
\end{equation}
This frequency corresponds to a chiral magnetic plasmon or, equivalently, the CMW in the strong-field limit. As we see,
the background magnetic field $B_0$ is responsible for the generation of the plasmon gap,
\begin{equation}
\label{Wigner-C-omega-gap}
\Omega_{B_0}= \sqrt{\frac{2v_Fe^2|eB_0|}{\pi}}.
\end{equation}
We note that the above value of the gap agrees with the one obtained in the LLL approximation
by a different method in Ref.~\cite{Fukushima:2015wck}. This is also consistent with the mass of the
resonance-like photon state revealed in QED in a strong magnetic field that is realized in the kinematic
regime $m^2\ll k_3^2\ll |eB|$ \cite{Gusynin:1998zq}.

\begin{figure}[tbp]
\centering
\includegraphics[width=0.55\textwidth]{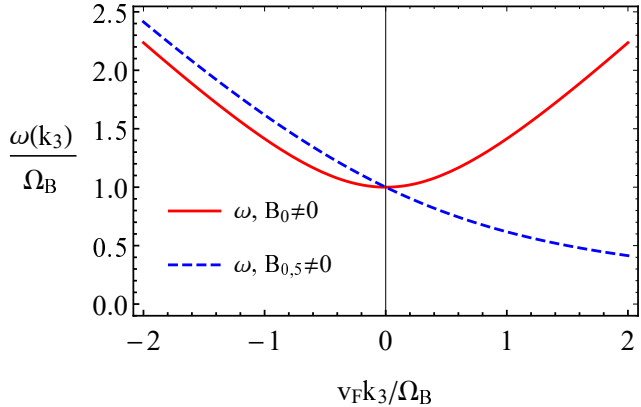}
\caption{\label{fig:Wigner-C-omega-B5}
The frequencies of the collective excitations in the presence of background magnetic (red solid line)
and axial magnetic (blue dashed line) fields, as given by Eq.~(\ref{Wigner-C-omega-B-sol}) and
Eq.~(\ref{Wigner-C-omega-B5-sol}), respectively. Here $\Omega_{B}=\Omega_{B_0}$ at $B_{0}\neq0$
and $\Omega_{B}=\Omega_{B_{0,5}}$ at $B_{0,5}\neq0$.}
\end{figure}

Further, let us study the case when there is only an axial magnetic field present, i.e., $B_{0}=0$ but $B_{0,5}\neq0$. In this case, the susceptibility  tensor reads
\begin{equation}
\label{Wigner-C-P3-B5}
\chi^{33} =  -\frac{v_Fe^2|eB_{0,5}|}{2\omega\pi^2(\omega+s_Bv_Fk_3)} .
\end{equation}
By making use of the characteristic equation (\ref{second-collective-B-tensor-dispersion-relation-general}),
we then derive the following dispersion relation of collective modes:
\begin{equation}
\label{Wigner-C-omega-B5-sol}
\omega=\frac{1}{2}\left(\sqrt{v_F^2k_3^2+\frac{8v_Fe^2|eB_{0,5}|}{\pi}}-s_Bv_Fk_3\right).
\end{equation}
By analogy with the CMW, we call this mode a chiral pseudomagnetic wave. The dispersion relations for
both types of collective modes (\ref{Wigner-C-omega-B-sol}) and (\ref{Wigner-C-omega-B5-sol})
are plotted in figure~\ref{fig:Wigner-C-omega-B5} as functions of the longitudinal wave vector $k_3$.
The results for a nonzero ordinary magnetic field $B_{0}$ (at $B_{0,5}=0$) and a nonzero axial
magnetic field $B_{0,5}$ (at $B_{0}=0$) are shown by solid red and blue dashed lines, respectively.
Note that, the asymmetry of the dispersion relation of the chiral pseudomagnetic wave is correlated
with the sign of $eB_{0,5}$. In figure~\ref{fig:Wigner-C-omega-B5}, we plotted the results for $eB_{0,5}>0$.
The results for $eB_{0,5}<0$ can be obtained simply by replacing $k_3\to-k_3$.

As we can see from figure~\ref{fig:Wigner-C-omega-B5}, the chiral pseudomagnetic wave,
which is realized in the case of a nonzero axial magnetic field $B_{0,5}\neq 0$, is qualitatively
different from the gapped chiral magnetic wave in the presence of a usual magnetic field. Indeed,
while the frequency of the chiral pseudomagnetic wave takes a nonzero value $\Omega_{B_{0,5}}
= \sqrt{2v_Fe^2 |eB_{0,5}|/\pi}$ at $k_3=0$, its dependence on the wave vector $k_3$ is asymmetric.
The corresponding mode appears to be gapless in the strong-field limit because $\omega\to 0$
at large positive $k_3$ when $eB_{0,5}>0$ (or large negative $k_3$ when $eB_{0,5}<0$). We argue
that the gaplessness of this mode is an artifact of the LLL approximation, which may be formally viewed
as the $B_{0,5}\to \infty$ limit. In the case of a strong but finite axial magnetic field, the inclusion of
higher Landau levels should make the corresponding mode gapped with the minimum of the energy
obtained at $k_3\sim B_{0,5}$. Indeed, this would be consistent with the weak-field analysis in
Refs.~\cite{Gorbar:2016ygi,Gorbar:2016sey}, where such an asymmetric dispersion relation with a minimum at
$k_3\sim B_{0,5}$ was predicted. The solution in Eq.~(\ref{Wigner-C-omega-B5-sol}) is nothing
else, but a strong-field version of the chiral pseudomagnetic wave that was first obtained in
Refs.~\cite{Gorbar:2016ygi,Gorbar:2016sey,Gorbar:2017cmw}.

\section{Thermoelectric phenomena in a strong magnetic field}
\label{sec:Wigner-C-LLL-Termal}

In this section, we study the thermoelectric transport in a chiral plasma in a strong magnetic field.
In essence, the problem reduces to determining the corrections to the Wigner function
in the LLL approximation and calculating the electric and heat (thermal) currents when an additional
weak electric field $\mathbf{E}$ and a small temperature gradient $\bm{\nabla}T \ne 0$ are present.
[We will take into account only the contribution due to charged chiral particles leaving aside all
other possible contributions.] For simplicity, we will assume that
$\mathbf{B}_{0,\lambda}\parallel\mathbf{E}\parallel\bm{\nabla}T\parallel\hat{\mathbf{z}}$.
In this case, the equations for the Wigner function components (\ref{Wigner-Weyl-Eq-eqs-be})
and (\ref{Wigner-Weyl-Eq-eqs-ee}) read
\begin{subequations}
\label{Wigner-C-LLL-T-eqs-all}
\begin{eqnarray}
\label{Wigner-C-LLL-T-eqs-be}
&&\partial_t w_0+v_F\left(\bm{\nabla}\cdot\mathbf{w}\right) +ev_F\left(\left[\mathbf{B}_{0,\lambda}\times\partial_{\mathbf{p}}\right]\cdot\mathbf{w}\right)
+ e(\mathbf{E}\cdot\partial_{\mathbf{p}})w_0^{(0)} =0,\\
&&\partial_t\mathbf{w}+v_F\bm{\nabla}w_0+ev_F\left[\mathbf{B}_{0,\lambda}\times\partial_{\mathbf{p}}\right]w_0  + 2\lambda v_F\left[\mathbf{p}\times\mathbf{w}\right] +e(\mathbf{E}\cdot\partial_{\mathbf{p}})\mathbf{w}^{(0)}=0.
\label{Wigner-C-LLL-T-eqs-ee}
\end{eqnarray}
\end{subequations}
Because of the spatial gradient of temperature, the Wigner function depends on spatial coordinates, i.e.,
$\bm{\nabla}W=(\bm{\nabla}T)\partial_TW\approx(\bm{\nabla}T)\partial_TW^{(0)}$. Here we assume that
the gradient $\bm{\nabla}T$ is small and of the same order of magnitude as $\mathbf{E}$.

By using the Wigner function components calculated in the LLL approximation, see
Eqs.~(\ref{Wigner-C-LLL-w0}) and (\ref{Wigner-C-LLL-wvec}), as well as introducing
a phenomenological collision term in the relaxation time approximation
on the right-hand sides of Eqs.~(\ref{Wigner-C-LLL-T-eqs-be}) and (\ref{Wigner-C-LLL-T-eqs-ee}),
we obtain the following set of equations:
\begin{subequations}
\label{Wigner-C-LLL-T-eqs1-all}
\begin{eqnarray}
\label{Wigner-C-LLL-T-eqs1-be}
e^{-p_{\perp}^2/|eB_{0,\lambda}|}\left[(-s_B\lambda)v_F (\nabla_3T) \partial_T +eE_3\partial_{p_3}\right] \tilde{f}_{LLL}(p_3) &=& -\frac{w_0 -e^{-p_{\perp}^2/|eB_{0,\lambda}|} \tilde{f}_{\rm LLL}(p_3)}{\tau},\\
e^{-p_{\perp}^2/|eB_{0,\lambda}|}\left[v_F (\nabla_3T) \partial_T +(-s_B\lambda)eE_3\partial_{p_3}\right] \tilde{f}_{LLL}(p_3) &=&  -\frac{w_3 +s_{B}\lambda e^{-p_{\perp}^2/|eB_{0,\lambda}|} \tilde{f}_{\rm LLL}(p_3)}{\tau},\nonumber\\
\label{Wigner-C-LLL-T-eqs1-ee}
\end{eqnarray}
\end{subequations}
where $\tau$ is the relaxation time. As is easy to check,  $w_1=w_2=0$ in the present setup. It is
worth noting that, in realistic models, the relaxation time may depend on the particle energy, external
fields, as well as other parameters. For simplicity, here we will assume that $\tau$ is a constant.
The solutions to Eqs.~(\ref{Wigner-C-LLL-T-eqs1-be}) and (\ref{Wigner-C-LLL-T-eqs1-ee}) read
\begin{subequations}
\label{Wigner-C-LLL-T-eqs-sol-all}
\begin{eqnarray}
\label{Wigner-C-LLL-T-eqs-sol-be}
w_0&=& e^{-p_{\perp}^2/|eB_{0,\lambda}|}\left\{\tilde{f}_{LLL}(p_3) -\tau\left[(-s_B\lambda)v_F \nabla_3T \partial_T\tilde{f}_{LLL}(p_3) +eE_3\partial_{p_3}\tilde{f}_{LLL}(p_3)\right] \right\},\\
w_3&=& -s_B\lambda e^{-p_{\perp}^2/|eB_{0,\lambda}|}\left\{\tilde{f}_{LLL}(p_3) -\tau\left[(-s_B\lambda)v_F \nabla_3T \partial_T\tilde{f}_{LLL}(p_3) +eE_3\partial_{p_3}\tilde{f}_{LLL}(p_3)\right] \right\}.\nonumber\\
\label{Wigner-C-LLL-T-eqs-sol-ee}
\end{eqnarray}
\end{subequations}
By making use of these results, we can now calculate the electric and heat current densities.
In terms of the Wigner function, the corresponding current densities are given by (see, e.g.,
Ref.~\cite{Lundgren:2014hra})
\begin{subequations}
\label{Wigner-C-LLL-T-je-jh-def}
\begin{eqnarray}
\label{Wigner-C-LLL-T-je-def}
j_3 &=& -ev_F\sum_{\lambda=\pm}\lambda \int \frac{d^3\mathbf{p}}{(2\pi)^3} \tr{\left[\sigma_3W\right]} =  -2ev_F \sum_{\lambda=\pm}\int \frac{d^3\mathbf{p}}{(2\pi)^3}  w_3,\\
j_3^{Q} &=& v_F\sum_{\lambda=\pm}\lambda \int \frac{d^3\mathbf{p}}{(2\pi)^3}  \,\left(\epsilon_{0}-\mu_{\lambda}\right)\tr{\left[\sigma_3W\right]} =  2v_F \sum_{\lambda=\pm}\int \frac{d^3\mathbf{p}}{(2\pi)^3}  \left(s_B\lambda v_Fp_3-\mu_{\lambda}\right)w_3.\nonumber\\
\label{Wigner-C-LLL-T-jh-def}
\end{eqnarray}
\end{subequations}
The calculation reduces to four types of integrals, presented in Eqs.~(\ref{Wigner-C-LLL-T-I1})
through (\ref{Wigner-C-LLL-T-I4}) in appendix~\ref{sec:App-Formulas}. By making use of the
corresponding results, we arrive at the following final expressions for the current densities:
\begin{subequations}
\label{Wigner-C-LLL-T-je-jh-1}
\begin{eqnarray}
\label{Wigner-C-LLL-T-je-1}
j_3 &=& \sum_{\lambda=\pm}2ev_Fs_B\lambda \left\{I_1-\tau\left[(-s_B\lambda)v_F(\nabla_3T)(\partial_TI_1)+eE_3I_3\right]\right\}\nonumber\\
&=&\sum_{\lambda=\pm}\frac{\lambda e^2B_{0,\lambda} \mu_{\lambda}}{(2\pi)^2} +\tau\sum_{\lambda=\pm}\frac{|eB_{0,\lambda}|e^2E_3v_F}{(2\pi)^2},\\
\label{Wigner-C-LLL-T-jh-1}
j_3^{Q} &=& -\sum_{\lambda=\pm}2v_Fs_B\lambda \left\{I_2-\mu_{\lambda}I_1-\tau\left[(-s_B\lambda)v_F(\nabla_3T)\partial_T(I_2-\mu_{\lambda}I_1)+eE_3(I_4-\mu_{\lambda}I_3)\right]\right\} \nonumber\\
&=&\sum_{\lambda=\pm}\frac{\lambda eB_{0,\lambda}}{2(2\pi)^2} \left[\mu_{\lambda}^2-\frac{\pi^2T^2}{3}+v_F^2\Lambda^2\right]
-\tau v_F(\nabla_3T)\sum_{\lambda=\pm} \frac{|eB_{0,\lambda}| T}{12}.
\end{eqnarray}
\end{subequations}
The last term in Eq.~(\ref{Wigner-C-LLL-T-je-1}) is similar to the usual Drude conductivity and is
related to the density of states on the LLL. Similar physical interpretation applies to the last term
in Eq.~(\ref{Wigner-C-LLL-T-jh-1}) connected with the heat transport. Note that the anomalous
Nernst effect, i.e., the correction to $\mathbf{j}$ proportional
to the cross product of the Berry curvature and the temperature gradient \cite{Xiao-Niu:2006},
is absent in the problem at hand. This fact is not surprising in view of the one-dimensional nature
of the LLL. Last but not least, we note that the last term in the square brackets in Eq.~(\ref{Wigner-C-LLL-T-jh-1}) diverges. The corresponding contribution should be naturally regularized in
realistic lattice models.

By comparing Eqs.~(\ref{Wigner-C-LLL-T-je-1}) and (\ref{Wigner-C-LLL-T-jh-1}) with the general
linear response relation, i.e.,
\begin{subequations}
\label{Wigner-C-LLL-T-je-jh-gen}
\begin{eqnarray}
\label{Wigner-C-LLL-T-je-gen}
j_3 &=& \tilde{j}_3 + \sigma_{33}^{(ee)}E_3 +\sigma_{33}^{(eT)}(-\nabla_3T), \\
\label{Wigner-C-LLL-T-jh-gen}
j_3^{Q} &=& \tilde{j}_3^{Q}+ \sigma_{33}^{(Te)}E_3 +\sigma_{33}^{(TT)}(-\nabla_3T),
\end{eqnarray}
\end{subequations}
where $\tilde{j}_3$ and $\tilde{j}_3^{Q}$ denote the nondissipative parts of the currents proportional
to $eB_0$ and $eB_{0,5}$ that come from the first terms in Eqs.~(\ref{Wigner-C-LLL-T-je-1})
and (\ref{Wigner-C-LLL-T-jh-1}) [see also Eqs.~(\ref{Wigner-Weyl-CME-CSE-answer-be}) and
(\ref{Wigner-Weyl-CME-CSE-answer-ee})], we find that the off-diagonal terms of the
thermoelectric conductivity tensor vanish, i.e., $\sigma_{33}^{(eT)}=T\sigma_{33}^{(Te)}=0$.
Therefore, the Seebeck coefficient (or thermopower) $S=\sigma_{33}^{(eT)}/\sigma_{33}^{(ee)}$
vanishes too. The thermoconductivity is defined in the absence of electric current and equals to
\begin{equation}
\label{Wigner-C-LLL-T-thermal-cond}
\kappa_{33}=\sigma_{33}^{(TT)}-\frac{\sigma_{33}^{(Te)}\sigma_{33}^{(eT)}}{\sigma_{33}^{(ee)}}=\sigma_{33}^{(TT)}.
\end{equation}
Finally, one can easily check that the Wiedemann-Franz law $\kappa_{33}=T\sigma_{33}^{(ee)}\pi^2/(3e^2)$
and the Mott relation $\sigma_{33}^{(eT)}=\pi^2T (\partial_{\mu}\sigma_{33}^{(ee)})/(3e)=0$ are satisfied.
In fact, in the present setup, $\kappa_{33}$ and $\sigma_{33}^{(ee)}$ are the only nonzero
components of the thermal and electric conductivity tensors, respectively.

In the case of the vanishing axial magnetic field, $B_{0,5}=0$, Eqs.~(\ref{Wigner-C-LLL-T-je-1})
and (\ref{Wigner-C-LLL-T-jh-1}) take the following simpler form:
\begin{subequations}
\label{Wigner-C-LLL-T-je-jh-B}
\begin{eqnarray}
\label{Wigner-C-LLL-T-je-B}
j_3 &=& \frac{e^2B_{0} \mu_{5}}{2\pi^2} +\tau \frac{e^2E_3v_F|eB_{0}|}{2\pi^2},\\
\label{Wigner-C-LLL-T-jh-B}
j_3^{Q} &=&\frac{eB_0\mu\mu_5}{2\pi^2} - \tau v_F|eB_{0}| \frac{T\nabla_3T}{6}.
\end{eqnarray}
\end{subequations}
On the other hand, when the ordinary magnetic field vanishes,
but a nonzero axial magnetic field is present, the electric current density is given by
\begin{equation}
\label{Wigner-C-LLL-T-je-B5}
j_3 = \frac{e^2B_{0,5} \mu}{2\pi^2} +\tau \frac{e^2E_3v_F|eB_{0,5}|}{2\pi^2}.
\end{equation}
As we see, this is similar to that in Eq.~(\ref{Wigner-C-LLL-T-je-B}). However, this is not the case for the heat current density, i.e.,
\begin{equation}
\label{Wigner-C-LLL-T-jh-B5}
j_3^{Q} = \frac{eB_{0,5}}{(2\pi)^2}\left(\mu^2+\mu^2_5-\frac{\pi^2T^2}{3}+v_F^2\Lambda^2\right)
- \tau v_F|eB_{0,5}| \frac{T\nabla_3T}{6}.
\end{equation}
While the dissipative term proportional to the relaxation time is similar to the corresponding
term in Eq.~(\ref{Wigner-C-LLL-T-jh-B}), the nondissipative ones are completely different. For example,
they do not require the presence of both chemical and chiral chemical potentials and could be nonzero
even at $\mu=\mu_5=0$.

Before concluding this section, it is instructive to point that all nondissipative contributions
in Eqs.~(\ref{Wigner-C-LLL-T-je-B5}) and (\ref{Wigner-C-LLL-T-jh-B5}), as well as in
Eqs.~(\ref{Wigner-C-LLL-T-je-B}) and (\ref{Wigner-C-LLL-T-jh-B}), are bound currents
that can be expressed as curls of other quantities. (In other words, their structure is similar to the magnetization
current $\mathbf{j}_{M} \sim[\bm{\nabla} \times\mathbf{M}]$.) This follows from the fact
that both $\mathbf{B}_{0}$ and $\mathbf{B}_{0,5}$ can be expressed as curls of the vector
$\mathbf{A}$ and axial vector $\mathbf{A}_5$ potentials, respectively. It is interesting to note
here that, in the context of Weyl semimetals, the axial potential $\mathbf{A}_5$ (unlike the usual
vector potential $\mathbf{A}$) is an observable quantity
\cite{Cortijo:2016yph,Cortijo:2016,Grushin-Vishwanath:2016,Pikulin:2016,Liu-Pikulin:2016} which is related to the separation between Weyl nodes in the momentum space.

\section{Summary}
\label{sec:summary}

By using the exact solutions of the Weyl equation for chiral fermions in constant magnetic and axial magnetic fields,
we calculated the equal-time Wigner function for a magnetized chiral fermion plasma at finite chemical potential
and temperature. This exact Wigner function is defined by the scalar
and vector parts in the basis of the Pauli matrices. While the vector part is necessary to calculate
currents, the scalar part defines the Wigner quasiprobability distribution function. We checked that,
to the linear order in magnetic field, the latter also agrees with the distribution functions of the chiral
kinetic theory. It is interesting to note that, owing to its quantum nature, the scalar part of the Wigner
function can be negative in a magnetic field. Besides the possibility of negative values (assuming
$\mu_\lambda>0$), the most crucial difference between the standard quasiprobability function,
which is given in terms of the Fermi-Dirac functions, and the Wigner quasiprobability distribution
function in a magnetic field is connected with the chirality-correlated asymmetric dependence of the latter on the longitudinal component of momentum.
A similar asymmetric dependence is found in the vector part of the Wigner function and is
principal for reproducing the correct chiral magnetic and chiral separation effects.

Retaining only the lowest Landau level contribution, the equation for the equal-time Wigner function
in a strong magnetic field is obtained. The constant background magnetic and axial magnetic (or, equivalently,
strain-induced pseudomagnetic) fields are taken into account nonperturbatively. In this case the scalar
and vector parts of the Wigner function are both proportional to the distribution function on the LLL.
By making use of this equation, it is found that the longitudinal collective excitations in a strong magnetic
field are gapped plasmons. The magnitude of their energy gap is determined by the value of the
magnetic field. Interestingly, the situation changes qualitatively in the case of the axial magnetic field.
The dispersion relation of the corresponding collective excitation, identified as the chiral pseudomagnetic
wave, is clearly asymmetric in the wave vector. While the chiral pseudomagnetic
wave appears to be gapless in the LLL approximation, we argued that the corresponding
mode is in fact gapped when higher Landau levels are included. As in the limit of a weak axial
magnetic field \cite{Gorbar:2016ygi,Gorbar:2016sey}, the minimum energy of the corresponding mode should be at $k_3\sim B_{0,5}$.

By making use of the Wigner function in the LLL approximation, we also studied the thermoelectric
transport of chiral fermions in a strong magnetic field. The analysis was performed in a phenomenological
model where the effects of collisions were introduced into the equation for the Wigner function
via a constant relaxation time. The latter, of course, is not a very realistic approximation to capture all details of the thermoelectric transport, but should be sufficient at least for understanding qualitative features.
We found that the electric and heat (thermal) current densities are
determined by the density of states on the lowest Landau level. While the nondissipative part
of the heat current density in a magnetic field requires the presence of both chemical and chiral chemical
potentials, its counterpart in an axial magnetic field is nontrivial when at least one of these
parameters or temperature is present. All nondissipative contributions to currents come in the form
of bound currents that are curls of other quantities. The structures of the dissipative terms are similar
(up to the interchange $B_{0,5}\leftrightarrow B_{0}$) in the cases of background magnetic $B_{0}$
and axial magnetic $B_{0,5}$ fields.

\acknowledgments

The work of E.V.G. was supported partially by the Ukrainian State Foundation for Fundamental Research.
The work of V.A.M. and P.O.S. was supported by the Natural Sciences and Engineering Research Council of Canada.
The work of I.A.S. was supported in part by the U.S. National Science Foundation under Grant No.~PHY-1404232.

\paragraph{Note added.} When finishing this paper, we became aware of a partially overlapping study by
Xin-li Sheng, Dirk H. Rischke, David Vasak, and Qun Wang \cite{SRVW}.

\appendix

\section{Wave functions of the Weyl Hamiltonian}
\label{sec:App-model-Weyl}

In this appendix, we derive the wave functions of the model Hamiltonian (\ref{model-Hamiltonian})
for the Weyl fermions in a constant magnetic field. In order to solve the eigenvalue problem
$H_{\lambda}\psi_{\lambda}=E\psi_{\lambda}$, we look for a solution in the form
$\psi_{\lambda}(\mathbf{x})=e^{ip_3x_3 +ip_2x_2 -iEt}\varphi_{\lambda}(\xi)$, where
the new variable $\xi$ is defined by
\begin{equation}
\xi=\sqrt{|eB_{\lambda}|}\left(\frac{p_2}{eB_{\lambda}}+x_1 \right).
\label{App-model-Bn0-xi}
\end{equation}
By noting that $\partial_{x_1}=\partial_{\xi}\sqrt{|eB_{\lambda}|}$, we check that function
$\varphi_{\lambda}$ satisfies the following ordinary differential equation:
\begin{equation}
\left[\partial_{\xi} -s_B\sigma_z \xi - \frac{i\lambda \sigma_x}{v_F \sqrt{|eB_{\lambda}|}} \left(E-\mu_{\lambda} -\lambda v_F p_3\sigma_z\right)\right]\varphi_{\lambda}=0,
\label{model-Bn0-Eq-2}
\end{equation}
where $s_{B}=\sign{eB_{\lambda}}$. The solution to this equation can be given in the
form
\begin{equation}
\varphi_{\lambda} = \Phi^{+}_{\lambda}(\xi)u^{+}_{\lambda}+\Phi^{-}_{\lambda}(\xi)u^{-}_{\lambda},
\label{model-Bn0-Phi}
\end{equation}
where $u^{\pm}_{\lambda}$ are two linearly independent spinors that satisfy the following relations:
\begin{equation}
\sigma_z u^{\pm}_{\lambda}=\pm u^{\pm}_{\lambda}, \qquad
u^{\mp}_{\lambda} = \frac{\lambda \sigma_x \left(E-\mu_{\lambda} -\lambda v_F p_3\sigma_z\right)}{ \sqrt{\left(E-\mu_{\lambda}\right)^2-v_F^2p_3^2}}u^{\pm}_{\lambda}.
\label{model-Bn0-u-pm}
\end{equation}
After substituting the ansatz (\ref{model-Bn0-Phi}) into Eq.~(\ref{model-Bn0-Eq-2}) and separating
linearly independent terms proportional to spinors $u_{\lambda}^{\pm}$, we arrive at the following
coupled set of equations:
\begin{eqnarray}
\label{model-Bn0-Eq-3}
&&\left[\partial_{\xi}\Phi^{\pm}_{\lambda}(\xi)\mp s_{B}\xi\Phi^{\pm}_{\lambda}(\xi)-ik\Phi^{\mp}_{\lambda}(\xi)\right]=0, \\
&&\left[\partial_{\xi}^2 \mp s_{B}-\xi^2+k^2\right]\Phi^{\pm}_{\lambda}(\xi)=0,
\label{model-Bn0-Eq-4}
\end{eqnarray}
where
\begin{equation}
k \equiv \sqrt{\frac{\left(E-\mu_{\lambda}\right)^2-v_F^2p_3^2}{v_F^2|eB_{\lambda}|}}.
\label{model-Bn0-k}
\end{equation}
Solutions of Eqs.~(\ref{model-Bn0-Eq-3}) and (\ref{model-Bn0-Eq-4}) can be expressed in
terms of the parabolic cylinder functions $D_{p}(\alpha \xi)$ \cite{Bateman,Gradshtein}, i.e.,
\begin{equation}
\Phi^{-s_B}_{\lambda}(\xi) = D_{k^2/2}\left(\sqrt{2}\xi\right), \qquad \Phi^{s_B}_{\lambda}(\xi) = -\frac{ik}{\sqrt{2}}D_{k^2/2-1}\left(\sqrt{2}\xi\right).
\label{model-Bn0-s-plus}
\end{equation}
By making use of Eq.~(\ref{model-Bn0-u-pm}), we also determine the explicit form of
spinors $u_{\lambda}^{\pm}$ for each of the two possible choices of $s_B=\pm1$. The
result reads
\begin{eqnarray}
u^{s_B}_{\lambda} &=&\mathcal{P}_{s_B}\left(
                         \begin{array}{c}
                           \lambda\frac{E-\mu_{\lambda} +\lambda s_B v_F p_3}{v_F\sqrt{2n|eB_{\lambda}|}} \\
                           0 \\
                         \end{array}
                       \right), \\
u^{-s_B}_{\lambda} &=& \mathcal{P}_{s_B}\left(
                        \begin{array}{c}
                          0 \\
                          1 \\
                        \end{array}
                      \right),
\label{model-Bn0-u-pm-sB=+1}
\end{eqnarray}
where
\begin{equation}
\mathcal{P}_{s_B} = \frac{(1-s_B)}{2} \sigma_x + \frac{(1+s_B)}{2}
\label{App-model-Bn0-PsB}
\end{equation}
is a matrix operator that interchanges the two components of the spinor when the sign
$s_{B}=\sign{eB_{\lambda}}$ changes.

The requirement of finite wave functions at $|\xi|\rightarrow\infty$ leads to the
constraint $k^2/2=n$, where $n=0, 1, 2 \ldots$. Then, the parabolic cylinder functions
can be expressed in terms of the Hermitian polynomials \cite{Bateman,Gradshtein}.
After fixing the overall normalization constants (by using formula 7.374.1 in Ref.~\cite{Gradshtein}),
we finally obtain the following eigenfunctions of the Weyl Hamiltonian (\ref{model-Hamiltonian}):
\begin{eqnarray}
\label{model-Bn0-WF-fin-n=0}
\psi_{n=0, p_2, p_3} &=& |eB_{\lambda}|^{1/4} e^{ip_3x_3 +ip_2x_2 -iEt}Y_0(\xi) \mathcal{P}_{s_B} \left(
                                                        \begin{array}{c}
                                                          0 \\
                                                          1 \\
                                                        \end{array}
                                                      \right),\\
\psi_{n>0, p_2, p_3} &=& |eB_{\lambda}|^{1/4} \sqrt{\frac{2v_F^2 n |eB_{\lambda}|}{2v_F^2 n |eB_{\lambda}|+\left[E_n-\mu_{\lambda}+s_{B}v_F\lambda p_z\right]^2}} e^{ip_3x_3 +ip_2x_2 -iEt} \mathcal{P}_{s_B}
\nonumber\\
&\times& \Bigg\{ Y_n(\xi)\left(
                                                        \begin{array}{c}
                                                          0 \\
                                                          1 \\
                                                        \end{array}
                                                      \right)
                                                      - i\lambda\frac{\left(E_n-\mu_{\lambda}+s_{B}v_F\lambda p_z\right)}{v_F\sqrt{2n|eB_{\lambda}|}}Y_{n-1}(\xi)\left(
                                                        \begin{array}{c}
                                                          1 \\
                                                          0 \\
                                                        \end{array}
                                                      \right)\Bigg\},
\label{model-Bn0-WF-fin-n>0}
\end{eqnarray}
where
\begin{equation}
Y_{n}(\xi) = \frac{1}{\sqrt{2^n n! \sqrt{\pi}}}e^{-\xi^2/2}H_{n}(\xi).
\label{App-model-Bn0-Yn}
\end{equation}
The corresponding energies for the lowest ($n=0$) and higher ($n>0$) Landau levels are
\begin{eqnarray}
\label{App-model-Bn0-E-be}
E_{n=0} &=& \mu_{\lambda}-s_{B}v_F\lambda p_3 = \mu_{\lambda} + \epsilon_0,\\
E_{n>0} &=& \mu_{\lambda} \pm v_F\sqrt{p_3^2+2n |eB_{\lambda}|} = \mu_{\lambda} + \epsilon_{n>0},
\label{App-model-Bn0-E-ee}
\end{eqnarray}
respectively. [Note that in the main text we changed the sign at $p_3$ in order to use the same notations in the Wigner function, where momenta are opposite with respect to that in the wave functions.]

\section{Derivation of the Wigner function in a constant magnetic field}
\label{sec:App-Wigner-der}

In this appendix, we provide the details of the calculation of the equal-time Wigner function of chiral fermions in a constant external magnetic field.
Let us write the Wigner operator (\ref{Wigner-Weyl-W-def}) explicitly
\begin{eqnarray}
\hat{W}_{\alpha\eta}(\mathbf{x},\mathbf{p}) &=& \frac{1}{2}\sum_{n,n^{\prime}}\int d^3 \mathbf{y} \int\frac{d^2\mathbf{q}}{(2\pi)^2}\int\frac{d^2 \mathbf{q}^{\prime}}{(2\pi)^2}e^{-i \mathbf{p}\cdot\mathbf{y}} e^{i\Phi (\mathbf{r}_{+},\mathbf{r}_{-})} \nonumber\\
&\times& \Bigg\{\left[2\hat{a}^{\dag}_{n,\mathbf{q}}\hat{a}_{n^{\prime},\mathbf{q}^{\prime}} -(2\pi)^2\delta_{n,n^{\prime}}\delta(\mathbf{q}-\mathbf{q}^{\prime})\right] \psi^{\dag}_{\eta, n,\mathbf{q}}(\mathbf{r}_{+})\psi_{\alpha, n^{\prime},\mathbf{q}^{\prime}}(\mathbf{r}_{-}) \nonumber\\
&+&\left[\hat{a}^{\dag}_{n,\mathbf{q}}\hat{b}^{\dag}_{n^{\prime},\mathbf{q}^{\prime}} -\hat{b}^{\dag}_{n^{\prime},\mathbf{q}^{\prime}}\hat{a}^{\dag}_{n,\mathbf{q}}\right] \psi^{\dag}_{\eta, n,\mathbf{q}}(\mathbf{r}_{+})\phi_{\alpha, n^{\prime},\mathbf{q}^{\prime}}(\mathbf{r}_{-})
\nonumber\\
&+& \left[\hat{b}_{n,\mathbf{q}}\hat{a}_{n^{\prime},\mathbf{q}^{\prime}} -\hat{a}_{n^{\prime},\mathbf{q}^{\prime}}\hat{b}_{n,\mathbf{q}}\right]\phi^{\dag}_{\eta, n,\mathbf{q}}(\mathbf{r}_{+})\psi_{\alpha, n^{\prime},\mathbf{q}^{\prime}}(\mathbf{r}_{-}) \nonumber\\
&-&\left[2\hat{b}^{\dag}_{n,\mathbf{q}}\hat{b}_{n^{\prime},\mathbf{q}^{\prime}} -(2\pi)^2\delta_{n,n^{\prime}}\delta(\mathbf{q}-\mathbf{q}^{\prime})\right] \phi^{\dag}_{\eta, n,\mathbf{q}}(\mathbf{r}_{+})\phi_{\alpha, n^{\prime},\mathbf{q}^{\prime}}(\mathbf{r}_{-}) \Bigg\},
\label{Wigner-Weyl-W-1}
\end{eqnarray}
where $\mathbf{r}_{\pm}=\mathbf{x}\pm \mathbf{y}/2$, the phase $\Phi(\mathbf{r}_{+},\mathbf{r}_{-})=-e\int_{\mathbf{r}_{-}}^{\mathbf{r}_{+}} d^3\mathbf{\mathbf{r}}\mathbf{A}_{\lambda}(\mathbf{\mathbf{r}}) = -eB_{\lambda}y_2x_1$ ensures the gauge
invariance of $\hat{W}_{\alpha\eta}$, and we used the standard anticommutation relations for the fermion particle creation and annihilation operators $\hat{a}^{\dag}_{n,\mathbf{q}}$, $\hat{a}_{n,\mathbf{q}}$, as well as their antiparticle $\hat{b}^{\dag}_{n,\mathbf{q}}$, $\hat{b}_{n,\mathbf{q}}$ counterparts. While the spinors for
particle states $\psi_{n,\mathbf{q}}$ are given by
Eqs.~(\ref{model-Bn0-WF-fin-n=0}) and (\ref{model-Bn0-WF-fin-n>0}), the spinors for antiparticles
are defined by $\left.\phi_{n,\mathbf{q}}\equiv\psi_{n,\mathbf{q}}\right|_{\epsilon_n \to -|\epsilon_n|}$. For simplicity,
we set also $\mathbf{q}=\left(q_2,q_3\right)$ and $\mathbf{q}^{\prime}=\left(q_2^{\prime},q_3^{\prime}\right)$.

The Wigner function is defined as an average of the Wigner operator over the Hilbert space of the multi-particle states
$|\Phi \rangle = | \ldots , N_{m_i}, \ldots , \ldots , \bar{N}_{m_i}, \ldots \rangle $
with $N_{m_i}$ particles in state $m_i$, and $\bar{N}_{m_i}$ antiparticles in state $m_i$, i.e.,
\begin{eqnarray}
W_{\alpha\eta}(\mathbf{x},\mathbf{p})&=&\mbox{Tr}\left(\hat{W}_{\alpha\eta}(\mathbf{x},\mathbf{p}) \hat{\rho}\right)= \sum_{\Phi} \langle \Phi | \hat{W}(\mathbf{x},\mathbf{p}) \hat{\rho} |\Phi\rangle
\nonumber\\
&=& \sum_{n=0}^{\infty}\int d^3 \mathbf{y} \int\frac{d^2\mathbf{q}}{(2\pi)^2}e^{-i \mathbf{p}\cdot\mathbf{y}-ieB_{\lambda}y_2x_1} 
\sum_{\epsilon_n} \mbox{tr}\Bigg\{\psi^{\dag}_{\eta, n,\mathbf{q}}(\mathbf{r}_{+})\psi_{\alpha, n,\mathbf{q}}(\mathbf{r}_{-}) \frac{\theta(\epsilon_{n}(\mathbf{q}))}{1+e^{\beta (\epsilon_{n}(\mathbf{q})-\mu_{\lambda})}} \nonumber\\
&-&\phi^{\dag}_{\eta, n,\mathbf{q}}(\mathbf{r}_{+})\phi_{\alpha, n,\mathbf{q}}(\mathbf{r}_{-}) \frac{\theta(-\epsilon_{n}(\mathbf{q}))}{1+e^{-\beta (\epsilon_{n}(\mathbf{q})-\mu_{\lambda})}} \nonumber\\
&-&\frac{1}{2}\left[\theta\left(\epsilon_n(\mathbf{k})\right)\psi^{\dag}_{\eta, n,\mathbf{q}}(\mathbf{r}_{+})\psi_{\alpha, n,\mathbf{q}}(\mathbf{r}_{-}) -\theta\left(-\epsilon_n(\mathbf{k})\right)\phi^{\dag}_{\eta, n,\mathbf{q}}(\mathbf{r}_{+})\phi_{\alpha, n,\mathbf{q}}(\mathbf{r}_{-})\right]\Bigg\}.
\label{Wigner-Weyl-W-Tr}
\end{eqnarray}
Here the density matrix operator $\hat{\rho}$ is given by Eq.~(\ref{Wigner-Weyl-rho-def}) in the main text.  Further, $\beta=1/T$ is the inverse
temperature, $\sum_{\epsilon_n}$ denotes the summation over positive and negative branches of the energy spectrum, and $\theta$-functions are the unit step
functions which select the proper sign of the energy of particles and antiparticles.

In order to proceed with the evaluation of the Wigner function (\ref{Wigner-Weyl-W-Tr}), we first calculate
\begin{eqnarray}
&&\int d^3 \mathbf{y} e^{-i \mathbf{p}\cdot\mathbf{y}} e^{-ieB_{\lambda}y_2x_1} \psi^{\dag}_{\eta, n,\mathbf{q}}\psi_{\alpha, n,\mathbf{q}} =\int d^3 \mathbf{y} e^{-i \mathbf{p}\cdot\mathbf{y}}
e^{-ieB_{\lambda}y_2x_1 -iq_3y_3-iq_2y_2} \nonumber\\
&&\times
\frac{\sqrt{|eB_{\lambda}|}}{1+\frac{[\epsilon_{n}(q_3)+s_Bv_F\lambda q_3]^2}{2v_F^2n|eB_{\lambda}|}}
\Bigg\{Y_n(\xi_{+})Y_n(\xi_{-}) \left[\frac{1+s_B}{2}\delta_{\alpha2}\delta_{\eta2} +\frac{1-s_B}{2}\delta_{\alpha1}\delta_{\eta1}\right] \nonumber\\
&&+\frac{[\epsilon_{n}(q_3)+s_Bv_F\lambda q_3]^2}{2nv_F^2|eB_{\lambda}|}Y_{n-1}(\xi_{+})Y_{n-1}(\xi_{-}) \left[\frac{1-s_B}{2}\delta_{\alpha2}\delta_{\eta2} +\frac{1+s_B}{2}\delta_{\alpha1}\delta_{\eta1}\right] \nonumber\\
&&-Y_n(\xi_{+})Y_{n-1}(\xi_{-})\frac{i\lambda[\epsilon_n(q_3)+s_B\lambda v_Fq_3]}{v_F\sqrt{2n|eB_{\lambda}|}} \left[\frac{1-s_B}{2}\delta_{\alpha2}\delta_{\eta1} +\frac{1+s_B}{2}\delta_{\alpha1}\delta_{\eta2}\right]\nonumber\\
&&+Y_{n-1}(\xi_{+})Y_{n}(\xi_{-})\frac{i\lambda[\epsilon_n(q_3)+s_B\lambda v_Fq_3]}{v_F\sqrt{2n|eB_{\lambda}|}} \left[\frac{1-s_B}{2}\delta_{\alpha1}\delta_{\eta2} +\frac{1+s_B}{2}\delta_{\alpha2}\delta_{\eta1}\right]\Bigg\},
\label{Wigner-Weyl-W-term}
\end{eqnarray}
where
\begin{eqnarray}
\label{Wigner-Weyl-xi}
\xi_{\pm} = \sqrt{|eB_{\lambda}|} \left(\frac{q_2}{eB_{\lambda}} +x_1 \pm \frac{y_1}{2}\right),\quad
\xi =  \sqrt{|eB_{\lambda}|} \left(\frac{q_2}{eB_{\lambda}} +x_1\right).
\end{eqnarray}

Using the table integral 7.377 in Ref.~\cite{Gradshtein}, we find the following auxiliary expressions:
\begin{eqnarray}
&&\int d y_1 e^{-ip_1y_1} e^{-\frac{|eB_{\lambda}|y_1^2}{4}}H_{n}\left[\sqrt{|eB_{\lambda}|} \left(\frac{q_2}{eB_{\lambda}} +x_1 + \frac{y_1}{2}\right)\right]H_{n}\left[\sqrt{|eB_{\lambda}|} \left(\frac{q_2}{eB_{\lambda}} +x_1 - \frac{y_1}{2}\right)\right] \nonumber\\
&&= (-1)^{n}\frac{e^{-p_1^2/|eB_{\lambda}|}}{\sqrt{|eB_{\lambda}|}} 2^{n+1}\sqrt{\pi} n! L_n \left[2\left(\xi^2+\frac{p_1^2}{|eB_{\lambda}|}\right)\right],
\label{Wigner-Weyl-W-int-y1}
\end{eqnarray}
and
\begin{eqnarray}
&&\int d y_1 e^{-ip_1y_1} e^{-\frac{|eB_{\lambda}|y_1^2}{4}}H_{n}\left[\sqrt{|eB_{\lambda}|} \left(\frac{q_2}{eB_{\lambda}} +x_1 \pm \frac{y_1}{2}\right)\right]H_{n-1}\left[\sqrt{|eB_{\lambda}|} \left(\frac{q_2}{eB_{\lambda}} +x_1 \mp \frac{y_1}{2}\right)\right] \nonumber\\
&&= (-1)^{n-1}\frac{e^{-p_1^2/|eB_{\lambda}|}}{\sqrt{|eB_{\lambda}|}} 2^{n+1}\sqrt{\pi} (n-1)! \left(\xi-\frac{ip_1}{\sqrt{|eB_{\lambda}|}}\right) L_{n-1}^1 \left[2\left(\xi^2+\frac{p_1^2}{|eB_{\lambda}|}\right)\right],
\label{Wigner-Weyl-W-int-y1-offdiag}
\end{eqnarray}
where $L_n^{m}(x)$ are the generalized Laguerre polynomials \cite{Gradshtein}. The above expressions allows us to obtain the diagonal
\begin{eqnarray}
&&\int d^3 \mathbf{y} e^{-i \mathbf{p}\cdot\mathbf{y}} e^{-ieB_{\lambda}y_2x_1} \psi^{\dag}_{\alpha, n,\mathbf{q}}\psi_{\alpha, n,\mathbf{q}}
=\frac{1}{2\pi} \delta(p_3+q_3) \delta(p_2+q_2+eB_{\lambda}x_1) e^{-\xi^2-p_1^2/|eB_{\lambda}|}  \nonumber\\
&&\times \frac{4(-1)^{n}v_F^2n|eB_{\lambda}|}{2v_F^2n|eB_{\lambda}|+[\epsilon_{n}(q_3)+s_Bv_F\lambda q_3]^2}
\Bigg\{L_n \left[2\left(\xi^2+\frac{p_1^2}{|eB_{\lambda}|}\right)\right] \left[\frac{1+s_B}{2}\delta_{\alpha2} +\frac{1-s_B}{2}\delta_{\alpha1}\right] \nonumber\\
&&-\frac{[\epsilon_{n}(q_3)+s_Bv_F\lambda q_3]^2}{2v_F^2n|eB_{\lambda}|}L_{n-1} \left[2\left(\xi^2+\frac{p_1^2}{|eB_{\lambda}|}\right)\right] \left[\frac{1-s_B}{2}\delta_{\alpha2} +\frac{1+s_B}{2}\delta_{\alpha1}\right] \Bigg\},
\label{Wigner-Weyl-W-int-2}
\end{eqnarray}
and off-diagonal $\alpha\neq\eta$
\begin{eqnarray}
&&\int d^3 \mathbf{y} e^{-i\mathbf{p}\cdot\mathbf{y}} e^{-ieB_{\lambda}y_2x_1} \psi^{\dag}_{\eta, n,\mathbf{q}}\psi_{\alpha, n,\mathbf{q}}
=\frac{1}{2\pi} \delta(p_3+q_3) \delta(p_2+q_2+eB_{\lambda}x_1)
 \frac{-i\lambda[\epsilon_n(q_3)+s_B\lambda v_Fq_3]}{v_Fn \sqrt{|eB_{\lambda}|}}
 \nonumber\\
&&\times
\frac{4(-1)^{n-1}v_F^2n|eB_{\lambda}|e^{-\xi^2-p_1^2/|eB_{\lambda}|}}{2v_F^2n|eB_{\lambda}| +[\epsilon_{n}(q_3)+s_Bv_F\lambda q_3]^2}
L_{n-1}^{1}\left[2\left(\xi^2+\frac{p_1^2}{|eB_{\lambda}|}\right)\right] \nonumber\\
&&\times
\Bigg\{\left(\frac{s_Bq_2-ip_1}{\sqrt{|eB_{\lambda}|}}+\sqrt{|eB_{\lambda}|}x_1\right) \left[\frac{1+s_B}{2}\delta_{\alpha1}\delta_{\eta2} +\frac{1-s_B}{2}\delta_{\alpha2}\delta_{\eta1}\right] \nonumber\\
&& -\left(\frac{s_Bq_2+ip_1}{\sqrt{|eB_{\lambda}|}}+\sqrt{|eB_{\lambda}|}x_1\right)\left[\frac{1-s_B}{2}\delta_{\alpha1}\delta_{\eta2} +\frac{1+s_B}{2}\delta_{\alpha2}\delta_{\eta1}\right] \Bigg\}.
\label{Wigner-Weyl-W-int-offdiag}
\end{eqnarray}
parts of the Wigner function $W_{\alpha \eta}(\mathbf{x},\mathbf{p})$. The latter can be represented in the following matrix form:
\begin{eqnarray}
W(\mathbf{x},\mathbf{p}) = \mathcal{P}_{s_B}\left(
      \begin{array}{cc}
        W_{11} & W_{12} \\
        W_{21} & W_{22} \\
      \end{array}
    \right)\mathcal{P}_{s_B}^{-1},
\label{Wigner-Weyl-W-ab}
\end{eqnarray}
where
\begin{eqnarray}
\label{Wigner-Weyl-W-11}
W_{11} &=&-e^{-p_{\perp}^2/|eB_{\lambda}|} \sum_{n=0}^{\infty} \sum_{\epsilon_n} \frac{(-1)^n L_{n-1}\left(\frac{2p_{\perp}^2}{|eB_{\lambda}|}\right)}{|\epsilon_n|}\Bigg\{ (|\epsilon_n|-s_B\lambda v_F p_3)\frac{\theta(\epsilon_n)}{1+e^{\beta(|\epsilon_n|-\mu_{\lambda})}}
\nonumber\\
&-&(|\epsilon_n|+s_B\lambda v_F p_3)\frac{\theta(-\epsilon_n)}{1+e^{\beta(|\epsilon_n|+\mu_{\lambda})}}
-\frac{1}{2} \left[\epsilon_n-s_B\lambda v_F p_3\right]
\Bigg\},\\
\label{Wigner-Weyl-W-22}
W_{22}&=&e^{-p_{\perp}^2/|eB_{\lambda}|} \sum_{n=0}^{\infty} \sum_{\epsilon_n} \frac{(-1)^n L_n\left(\frac{2p_{\perp}^2}{|eB_{\lambda}|}\right)}{|\epsilon_n|}\Bigg\{
(|\epsilon_n|+s_Bv_F\lambda p_3)\frac{\theta(\epsilon_n)}{1+e^{\beta(|\epsilon_n|-\mu_{\lambda})}}
\nonumber\\
&-&(|\epsilon_n|-s_Bv_F\lambda p_3)\frac{\theta(-\epsilon_n)}{1+e^{\beta(|\epsilon_n|+\mu_{\lambda})}}
-\frac{1}{2} \left[\epsilon_n+s_B\lambda v_F p_3\right]
\Bigg\},\\
\label{Wigner-Weyl-W-21}
W_{12}&=&W_{21}^{*}=2 e^{-p_{\perp}^2/|eB_{\lambda}|} \sum_{n=0}^{\infty} \sum_{\epsilon_n}(-1)^n \frac{\lambda v_F p_{-}}{|\epsilon_n|}L_{n-1}^{1}\left(\frac{2p_{\perp}^2}{|eB_{\lambda}|}\right) \nonumber\\
&\times&\Bigg\{\frac{\theta(\epsilon_n)}{1+e^{\beta(|\epsilon_n|-\mu_{\lambda})}}
+\frac{\theta(-\epsilon_n)}{1+e^{\beta(|\epsilon_n|+\mu_{\lambda})}} -\frac{1}{2}\Bigg\}.
\end{eqnarray}
Here $p_{\perp}^2=p_1^2+p_2^2$, $p_{\pm}=p_1\pm i s_Bp_2$, and we omitted the arguments of $W_{nm}$.
Then the Wigner function of form (\ref{Wigner-Weyl-w-Pauli-def}) can be easily determined using the following
relations:
\begin{equation}
\label{Wigner-Weyl-w-coef-be}
w_0\equiv\frac{W_{11}+W_{22}}{2},\quad
w_1\equiv\lambda \frac{W_{12}+W_{21}}{2},\quad
w_2\equiv i\lambda s_B\frac{W_{12}-W_{21}}{2},\quad
w_3\equiv\lambda s_B\frac{W_{11}-W_{22}}{2}.
\end{equation}

\section{Weak magnetic field limit}
\label{sec:App-Wigner-weak-B}

In this appendix, we derive the Wigner function in the limit of a weak magnetic field. The
coefficients $W_{11}$, $W_{22}$, $W_{12}$, and $W_{21}$ in Eqs.~(\ref{Wigner-Weyl-W-11}) through (\ref{Wigner-Weyl-W-21}) are
\begin{eqnarray}
\label{Wigner-Weyl-weak-B-W11}
W_{11}&\equiv& A_{-} -A_{+} +\delta A,\\
\label{Wigner-Weyl-weak-B-W22}
W_{22}&\equiv& B_{-} -B_{+} +\delta B,\\
\label{Wigner-Weyl-weak-B-W12}
W_{12}&\equiv& C_{-} +C_{+} +\delta C,\\
\label{Wigner-Weyl-weak-B-W21}
W_{21}&\equiv& W_{12}^{*},
\end{eqnarray}
where
\begin{eqnarray}
\label{Wigner-Weyl-weak-B-A-mp}
A_{\mp}&=& -e^{-p_{\perp}^2/|eB_{\lambda}|}  \sum_{n=0}^{\infty} (-1)^n L_{n-1}\left(\frac{2p_{\perp}^2}{|eB_{\lambda}|}\right) \frac{(|\epsilon_n|\mp s_B\lambda v_F p_3)}{|\epsilon_n|}\frac{\theta(\pm\epsilon_n)}{1+e^{\beta(|\epsilon_n|\mp\mu_{\lambda})}},\\
\label{Wigner-Weyl-weak-B-B-mp}
B_{\mp}&=& e^{-p_{\perp}^2/|eB_{\lambda}|} \sum_{n=0}^{\infty} (-1)^n L_n\left(\frac{2p_{\perp}^2}{|eB_{\lambda}|}\right) \frac{(|\epsilon_n|\pm s_B\lambda v_F p_3)}{|\epsilon_n|}\frac{\theta(\pm\epsilon_n)}{1+e^{\beta(|\epsilon_n|\mp\mu_{\lambda})}},\\
\label{Wigner-Weyl-weak-B-C-mp}
C_{\mp}&=&2 e^{-p_{\perp}^2/|eB_{\lambda}|} \sum_{n=0}^{\infty} (-1)^n L_{n-1}^{1}\left(\frac{2p_{\perp}^2}{|eB_{\lambda}|}\right) \frac{\lambda v_Fp_{-}}{|\epsilon_n|} \frac{\theta(\pm\epsilon_n)}{1+e^{\beta(|\epsilon_n|\mp\mu_{\lambda})}},
\end{eqnarray}
and
\begin{eqnarray}
\label{Wigner-C-weak-B-A}
\delta A&=& -e^{-p_{\perp}^2/|eB_{\lambda}|}  \sum_{n=0}^{\infty}  (-1)^n L_{n-1}\left(\frac{2p_{\perp}^2}{|eB_{\lambda}|}\right) \frac{s_B\lambda v_F p_3}{|\epsilon_n|},\\
\label{Wigner-C-weak-B-B}
\delta B&=&-e^{-p_{\perp}^2/|eB_{\lambda}|}  \sum_{n=0}^{\infty} (-1)^n L_n\left(\frac{2p_{\perp}^2}{|eB_{\lambda}|}\right) \frac{s_B\lambda v_F p_3}{|\epsilon_n|},\\
\label{Wigner-C-weak-B-C}
\delta C&=&-2e^{-p_{\perp}^2/|eB_{\lambda}|}  \sum_{n=0}^{\infty} (-1)^n \frac{\lambda v_F p_{-}}{|\epsilon_n|}L_{n-1}^{1}\left(\frac{2p_{\perp}^2}{|eB_{\lambda}|}\right).
\end{eqnarray}

In order to sum over all Landau levels, we employ the following tricks:
\begin{eqnarray}
\label{Wigner-Weyl-weak-B-A-expansion}
\frac{|\epsilon_n|\mp s_B\lambda v_F p_3}{|\epsilon_n|}\frac{\theta(\pm\epsilon_n)}{1+e^{\beta(|\epsilon_n|\mp\mu_{\lambda})}} &=& \sum_{m=0}^{\infty} (2n|eB_{\lambda}|)^{m} \frac{G_1^{(m)}(p_3^2)}{m!} \nonumber\\
&=& \lim_{s_1\to-0}\sum_{m=0}^{\infty} \left(\frac{d^{m}}{d s_1^{m}} e^{2s_1n|eB_{\lambda}|}\right) \frac{G_1^{(m)}(p_3^2)}{m!},
\\
\label{Wigner-Weyl-weak-B-B-expansion}
\frac{|\epsilon_n|\pm s_B\lambda v_F p_3}{|\epsilon_n|}\frac{\theta(\pm\epsilon_n)}{1+e^{\beta(|\epsilon_n|\mp\mu_{\lambda})}} &=&  \sum_{m=0}^{\infty} (2n|eB_{\lambda}|)^{m} \frac{G_2^{(m)}(p_3^2)}{m!} \nonumber\\
&=&\lim_{s_1\to-0}\sum_{m=0}^{\infty} \left(\frac{d^{m}}{d s_1^{m}} e^{2s_1n|eB_{\lambda}|}\right) \frac{G_2^{(m)}(p_3^2)}{m!},
\\
\label{Wigner-Weyl-weak-B-C-expansion}
\frac{1}{|\epsilon_n|}\frac{\theta(\pm\epsilon_n)}{1+e^{\beta(|\epsilon_n|\mp\mu_{\lambda})}} &=&  \sum_{m=0}^{\infty} (2n|eB_{\lambda}|)^{m}   \frac{G_3^{(m)}(p_3^2)}{m!}  \nonumber\\
&=&
\lim_{s_1\to-0}\sum_{m=0}^{\infty} \left(\frac{d^{m}}{d s_1^{m}} e^{2s_1n|eB_{\lambda}|}\right)  \frac{G_3^{(m)}(p_3^2)}{m!},\\
\label{Wigner-C-weak-expansion}
\frac{1}{|\epsilon_n|}= \sum_{m=0}^{\infty} (2n|eB_{\lambda}|)^{m} \frac{G_4^{(m)}(p_3^2)}{m!}
&=& \lim_{s_1\to-0}\sum_{m=0}^{\infty} \left(\frac{d^{m}}{d s_1^{m}} e^{2s_1n|eB_{\lambda}|}\right)  \frac{G_4^{(m)}(p_3^2)}{m!},
\end{eqnarray}
where $G_i^{(m)}$ denotes the m-th derivative with respect to its argument and the following shorthand notations are used:
\begin{eqnarray}
\label{Wigner-Weyl-weak-B-G1}
G_1\left(p_3^2+2n|eB_{\lambda}|\right)&=&\frac{|\epsilon_n|\mp s_B\lambda v_F p_3}{|\epsilon_n|}\frac{\theta(\pm\epsilon_n)}{1+e^{\beta(|\epsilon_n|\mp\mu_{\lambda})}},\\
\label{Wigner-Weyl-weak-B-G2}
G_2\left(p_3^2+2n|eB_{\lambda}|\right)&=&\frac{|\epsilon_n|\pm s_B\lambda v_F p_3}{|\epsilon_n|}\frac{\theta(\pm\epsilon_n)}{1+e^{\beta(|\epsilon_n|\mp\mu_{\lambda})}},\\
\label{Wigner-Weyl-weak-B-G3}
G_3\left(p_3^2+2n|eB_{\lambda}|\right)&=&\frac{1}{|\epsilon_n|}\frac{\theta(\pm\epsilon_n)}{1+e^{\beta(|\epsilon_n|\mp\mu_{\lambda})}},\\
\label{Wigner-C-weak-B-G4}
G_4\left(p_3^2+2n|eB_{\lambda}|\right)&=&\frac{1}{|\epsilon_n|}.
\end{eqnarray}

Performing the summation over Landau levels by using formula 7.414.8 in Ref.~\cite{Gradshtein}, we obtain
\begin{eqnarray}
A_{\mp} &=&e^{-p_{\perp}^2/|eB_{\lambda}|} \lim_{s_1\to-0}\sum_{m=0}^{\infty} \frac{d^{m}}{d s_1^{m}}\left[ e^{2s_1|eB_{\lambda}|} \frac{\exp{\left(\frac{p_{\perp}^2}{|eB_{\lambda}|\left(1+e^{-2s_1|eB_{\lambda}|}\right)}\right)}}{1+e^{2s_1|eB_{\lambda}|}} \frac{G_1^{(m)}(p_3^2)}{m!}\right]\nonumber\\
&\simeq& \frac{1}{2} \sum_{m=0}^{\infty} \frac{(p_{\perp}^2)^m}{m!} G_{1}^{(m)}(p_3^2) +\frac{|eB_{\lambda}|}{2}\sum_{m=0}^{\infty} \frac{(p_{\perp}^2)^{m-1}}{(m-1)!} G_{1}^{(m)}(p_3^2) +O(|eB_{\lambda}|^2)\nonumber\\
&\simeq&\frac{1}{2} G_1(\epsilon_{\mathbf{p}})
+\frac{|eB_{\lambda}|}{2} \left(\frac{d}{dp_{\perp}^2} G_1(\epsilon_{\mathbf{p}})\right)+O(|eB_{\lambda}|^2),
\label{Wigner-Weyl-weak-B-A-1}
\end{eqnarray}

\begin{eqnarray}
B_{\mp} &=&e^{-p_{\perp}^2/|eB_{\lambda}|} \lim_{s_1\to-0}\sum_{m=0}^{\infty} \frac{d^{m}}{d s_1^{m}}\left[ \frac{\exp{\left(\frac{p_{\perp}^2}{|eB_{\lambda}|\left(1+e^{-2s_1|eB_{\lambda}|}\right)}\right)}}{1+e^{2s_1|eB_{\lambda}|}} \frac{G_2^{(m)}(p_3^2)}{m!}\right]\nonumber\\
&\simeq& \frac{1}{2} \sum_{m=0}^{\infty} \frac{(p_{\perp}^2)^m}{m!} G_{2}^{(m)}(p_3^2) -\frac{|eB_{\lambda}|}{2}\sum_{m=0}^{\infty} \frac{(p_{\perp}^2)^{m-1}}{(m-1)!} G_{2}^{(m)}(p_3^2) +O(|eB_{\lambda}|^2)\nonumber\\
&\simeq&\frac{1}{2} G_2(\epsilon_{\mathbf{p}})
-\frac{|eB_{\lambda}|}{2} \left(\frac{d}{dp_{\perp}^2} G_2(\epsilon_{\mathbf{p}})\right)+O(|eB_{\lambda}|^2),
\label{Wigner-Weyl-weak-B-B-1}
\end{eqnarray}

\begin{eqnarray}
C_{\mp} &=& -2\lambda v_Fp_{-} e^{-p_{\perp}^2/|eB_{\lambda}|} \lim_{s_1\to-0}\sum_{m=0}^{\infty} \frac{d^{m}}{d s_1^{m}}\left[ e^{2s_1|eB_{\lambda}|} \frac{\exp{\left(\frac{p_{\perp}^2}{|eB_{\lambda}|\left(1+e^{-2s_1|eB_{\lambda}|}\right)}\right)}}{(1+e^{2s_1|eB_{\lambda}|})^2} \frac{G_3^{(m)}(p_3^2)}{m!}\right]\nonumber\\
&\simeq& -\frac{\lambda v_Fp_{-}}{2} \sum_{m=0}^{\infty} \frac{(p_{\perp}^2)^m}{m!} G_{3}^{(m)}(p_3^2) +O(|eB_{\lambda}|^2)
\simeq-\frac{\lambda v_Fp_{-}}{2} G_3(\epsilon_{\mathbf{p}}) +O(|eB_{\lambda}|^2),
\label{Wigner-Weyl-weak-B-C-1}
\end{eqnarray}
where $\epsilon_{\mathbf{p}}\equiv v_F|\mathbf{p}|$. The coefficients $\delta A$, $\delta B$, and $\delta C$ read
\begin{eqnarray}
\delta A &=&s_B\lambda v_F p_3 e^{-p_{\perp}^2/|eB_{\lambda}|} \lim_{s_1\to-0}\sum_{m=0}^{\infty} \frac{d^{m}}{d s_1^{m}}\left[ e^{2s_1|eB_{\lambda}|} \frac{\exp{\left(\frac{p_{\perp}^2}{|eB_{\lambda}|\left(1+e^{-2s_1|eB_{\lambda}|}\right)}\right)}}{1+e^{2s_1|eB_{\lambda}|}} \frac{G_4^{(m)}(p_3^2)}{m!}\right]\nonumber\\
&\simeq& \frac{s_B\lambda v_F p_3}{2} \sum_{m=0}^{\infty} \frac{(p_{\perp}^2)^m}{m!} G_{4}^{(m)}(p_3^2) +s_B\lambda v_F p_3\frac{|eB_{\lambda}|}{2}\sum_{m=0}^{\infty} \frac{(p_{\perp}^2)^{m-1}}{(m-1)!} G_{4}^{(m)}(p_3^2) +O(|eB_{\lambda}|^2)\nonumber\\
&\simeq&\frac{s_B\lambda v_F p_3}{2} G_4(\epsilon_{\mathbf{p}})
+s_B\lambda p_3\frac{|eB_{\lambda}|}{2} \left(\frac{d}{dp_{\perp}^2} G_4(\epsilon_{\mathbf{p}})\right)+O(|eB_{\lambda}|^2) \nonumber\\
&=&\frac{s_B\lambda v_Fp_3}{2\epsilon_{\mathbf{p}}}
-s_B\lambda v_F^3p_3\frac{|eB_{\lambda}|}{4\epsilon_{\mathbf{p}}^3} +O(|eB_{\lambda}|^2),
\label{Wigner-C-weak-A-1}
\end{eqnarray}
\begin{eqnarray}
\delta B &=&-s_B\lambda v_F p_3 e^{-p_{\perp}^2/|eB_{\lambda}|} \lim_{s_1\to-0}\sum_{m=0}^{\infty} \frac{d^{m}}{d s_1^{m}}\left[ \frac{\exp{\left(\frac{p_{\perp}^2}{|eB_{\lambda}|\left(1+e^{-2s_1|eB_{\lambda}|}\right)}\right)}}{1+e^{2s_1|eB_{\lambda}|}} \frac{G_4^{(m)}(p_3^2)}{m!}\right]\nonumber\\
&\simeq& \frac{-s_B\lambda v_F p_3}{2} \sum_{m=0}^{\infty} \frac{(p_{\perp}^2)^{m}}{m!} G_{4}^{(m)}(p_3^2) +s_B\lambda v_F p_3\frac{|eB_{\lambda}|}{2}\sum_{m=0}^{\infty} \frac{(p_{\perp}^2)^{m-1}}{(m-1)!} G_{4}^{(m)}(p_3^2) +O(|eB_{\lambda}|^2)\nonumber\\
&\simeq&-\frac{s_B\lambda v_F p_3}{2} G_4(\epsilon_{\mathbf{p}})
+s_B\lambda v_F p_3\frac{|eB_{\lambda}|}{2} \left(\frac{d}{dp_{\perp}^2} G_4(\epsilon_{\mathbf{p}})\right)+O(|eB_{\lambda}|^2)
\nonumber\\
&=&-\frac{s_B\lambda v_F p_3}{2\epsilon_{\mathbf{p}}}
-s_B\lambda v_F^3 p_3\frac{|eB_{\lambda}|}{4\epsilon_{\mathbf{p}}^3}+O(|eB_{\lambda}|^2),
\label{Wigner-C-weak-B-B-1}
\end{eqnarray}
\begin{eqnarray}
\delta C &=&-2\lambda v_Fp_{-} e^{-p_{\perp}^2/|eB_{\lambda}|} \lim_{s_1\to-0}(-1)\sum_{m=0}^{\infty} \frac{d^{m}}{d s_1^{m}}\left[ e^{2s_1|eB_{\lambda}|} \frac{\exp{\left(\frac{p_{\perp}^2}{|eB_{\lambda}|\left(1+e^{-2s_1|eB_{\lambda}|}\right)}\right)}}{(1+e^{2s_1|eB_{\lambda}|})^2} \frac{G_4^{(m)}(p_3^2)}{m!}\right]\nonumber\\
&\simeq& \frac{\lambda v_Fp_{-}}{2} \sum_{m=0}^{\infty} \frac{(p_{\perp}^2)^m}{m!} G_{3}^{(m)}(p_3^2) +O(|eB_{\lambda}|^2)
=\frac{\lambda v_Fp_{-}}{2} G_3(\epsilon_{\mathbf{p}}) +O(|eB_{\lambda}|^2)
\nonumber\\
&=&\frac{\lambda v_Fp_{-}}{2\epsilon_{\mathbf{p}}} +O(|eB_{\lambda}|^2).
\label{Wigner-C-weak-B-C-1}
\end{eqnarray}

Combining the above results together and using Eq.~(\ref{Wigner-Weyl-w-coef-be}), we obtain Eqs.~(\ref{Wigner-C-wB-w-be}) and
(\ref{Wigner-C-wB-w-ee}) in the main text.

\section{Useful formulas}
\label{sec:App-Formulas}

In this appendix, we present some key formulas used in the calculation of the electric and heat
current densities, defined by Eqs.~(\ref{Wigner-C-LLL-T-je-def}) and (\ref{Wigner-C-LLL-T-jh-def}) in
the main text.

Let us start by presenting the result for the following table integral:
\begin{eqnarray}
\int_0^{\infty} dp_3\, p_3^{n}\,  \frac{1}{1+e^{(v_Fp_3\mp \mu_{\lambda})/T}}
= -\frac{T^{n+1} \Gamma(n+1) }{v_F^{n+1}}  \mbox{Li}_{n+1}\left(-e^{\pm\mu_{\lambda}/T}\right),
\qquad n\geq 0,
\label{integral-3a}
\end{eqnarray}
where $\mbox{Li}_{n}(x)$ is the polylogarithm function (see formula 1.1.14 in Ref.~\cite{Erdelyi:Vol1}). [Note that in the given reference $\mathrm{Li}_n(x) \equiv \mathrm{F}(x, n)$.] The polylogarithm function at
$n=0,1$ can be rewritten as follows:
\begin{eqnarray}
\mbox{Li}_{0}\left(-e^{x}\right) &=& -\frac{1}{1+e^{-x}}, \\
\mbox{Li}_{1}\left(-e^{x}\right) &=& -\ln{\left(1+e^{x}\right)}.
\label{App-polylog}
\end{eqnarray}
The following identities for the polylogarithm functions are useful when taking into account the antiparticles contributions:
\begin{eqnarray}
\ln(1+e^{x})-\ln(1+e^{-x}) &=& x,\\
\mbox{Li}_{2} (-e^{x}) +\mbox{Li}_{2} (-e^{-x}) &=& -\frac{x^2}{2}-\frac{\pi^2}{6}.
\end{eqnarray}
By making use of the table integral in Eq.~(\ref{integral-3a}), it is straightforward to check the following
results for the four types of integrations encountered in the calculation of the electric and heat
current densities:
\begin{eqnarray}
\label{Wigner-C-LLL-T-I1}
I_1 &=& \int \frac{d^3\mathbf{p}}{(2\pi)^3} e^{-p_{\perp}^2/|eB_{0,\lambda}|}\tilde{f}_{LLL}(p_3)
=\frac{|eB_{0,\lambda}|}{2(2\pi)^2} \frac{\mu_{\lambda}}{v_F}, \\
\label{Wigner-C-LLL-T-I2}
I_2 &=& \int\frac{d^3\mathbf{p}}{(2\pi)^3} e^{-p_{\perp}^2/|eB_{0,\lambda}|}s_B\lambda v_Fp_3\tilde{f}_{LLL}(p_3)
=\frac{|eB_{0,\lambda}|}{4v_F(2\pi)^2} \left(\mu_{\lambda}^2+\frac{\pi^2T^2}{3} -v_F^2\Lambda^2\right),\\
\label{Wigner-C-LLL-T-I3}
I_3 &=& \int \frac{d^3\mathbf{p}}{(2\pi)^3} e^{-p_{\perp}^2/|eB_{0,\lambda}|} \partial_{p_3}\tilde{f}_{LLL}(p_3) =-\frac{eB_{0,\lambda}}{2(2\pi)^2} \lambda,\\
\label{Wigner-C-LLL-T-I4}
I_4&=& \int \frac{d^3\mathbf{p}}{(2\pi)^3} e^{-p_{\perp}^2/|eB_{0,\lambda}|} s_B\lambda v_Fp_3\,\partial_{p_3}\tilde{f}_{LLL}(p_3) = -\frac{eB_{0,\lambda}}{2(2\pi)^2} \lambda \mu_{\lambda},
\end{eqnarray}
where $\tilde{f}_{LLL}(p_3)$ is the Wigner function in the LLL approximation defined in Eq.~(\ref{Wigner-C-LLL-fLLL}).
It should be noted that the last term in the parentheses on the right-hand side of Eq.~(\ref{Wigner-C-LLL-T-I2})
contains a quadratic divergency that stems from $-\lambda s_B\sign{p_3}/2$ term in the function $\tilde{f}_{LLL}(p_3)$.

\end{document}